\documentclass[
  twocolumn,
  aps,
  prl,
  amsmath,
  amssymb,
  superscriptaddress,
  nofootinbib,
  floatfix
]{revtex4-1}
\usepackage{mathtools}
\usepackage{bm}
\usepackage{dsfont}
\usepackage{graphicx}
\usepackage{subfigure}
\usepackage{pifont}
\usepackage{textcomp}
\usepackage{gensymb}
\usepackage{accents}
\usepackage{upgreek}
\usepackage{array}
\usepackage{hyperref}
\usepackage{color}
\usepackage{lipsum}
\usepackage{tikz}

\usepackage{hyperref}
\hypersetup{colorlinks=true, citecolor=blue, urlcolor=blue, linkcolor=blue}

\makeatletter
\newcommand*{\supplementarystart}{%
  \close@column@grid%
  \clearpage%
  \onecolumngrid%
  \setcounter{enumiv}{0} 
  \setcounter{equation}{0} 
  \setcounter{figure}{0} 
  \setcounter{table}{0} 
  \setcounter{page}{1}
  \c@secnumdepth=4
  \renewcommand{\theequation}{s\arabic{equation}} 
  \renewcommand{\bibnumfmt}[1]{[s##1]} 
  \renewcommand{\cite}[1]{{[}\onlinecite{##1}{]}}
  \renewcommand{\thefigure}{s\arabic{figure}}
  \renewcommand{\thetable}{s\Roman{table}}
  \renewcommand{\thepage}{s\arabic{page}}
}
\makeatother

\newcommand{\be}{\begin{equation}}
\newcommand{\e}{\end{equation}}
\newcommand{\beml}{\begin{subequations}}
\newcommand{\eml}{\end{subequations}}
\newcommand{\beq}{\begin{eqnarray}}
\newcommand{\eq}{\end{eqnarray}}
\newcommand{\ba}{\begin{array}}
\newcommand{\ea}{\end{array}}
\newcommand{\bpm}{\begin{pmatrix}}
\newcommand{\epm}{\end{pmatrix}}
\newcommand{\bc}{\begin{cases}}
\newcommand{\ec}{\end{cases}}
\newcommand{\lt}{\left}
\newcommand{\rt}{\right}
\newcommand{\n}{\nonumber}
\newcommand{\la}{\langle}
\newcommand{\ra}{\rangle}

\newcommand{\bs}{\mathbf}
\newcommand{\bb}{\boldsymbol}
\newcommand{\h}{^\dagger}
\newcommand{\0}{^{\phantom{\dagger}}}

\include{filename}

\DeclareMathOperator{\dv}{div}

\begin{document}
	
\title{Optical detection of 4-spin chiral interaction in a 2D honeycomb ferrimagnet}	
\author{B. Murzaliev}
\affiliation{Radboud University, Institute for Molecules and Materials, NL-6525 AJ Nijmegen, the Netherlands}

\author{M. I. Katsnelson}
\affiliation{Radboud University, Institute for Molecules and Materials, NL-6525 AJ Nijmegen, the Netherlands}
\affiliation{Constructor Knowledge Institute, Constructor University,
28759 Bremen, Germany}

\author{M. Titov}
\affiliation{Radboud University, Institute for Molecules and Materials, NL-6525 AJ Nijmegen, the Netherlands}

\date{\today}

\begin{abstract}
The broken inversion symmetry of the magnetic lattice is normally described by Lifshitz invariants in a micromagnetic energy functional. Three exceptions are the lattices with T$_\textrm{d}$, C$_\textrm{3h}$ and D$_\textrm{3h}$ point group symmetries. For such lattices all Lifshitz invariants are forbidden despite broken inversion symmetry. In these cases the inversion symmetry breaking of the corresponding magnets is reflected by 4-spin chiral invariants that are not related to the Dzyaloshinskii-Moriya interaction. The experimental detection of 4-spin chiral interactions is an important task that has yet to be performed. We propose that the 4-spin chiral interaction can be probed by energy-selective magnon relaxation in two-dimensional ferromagnet Fe$_{3}$GeTe$_{2}$ that possess D$_\textrm{3h}$ point group symmetry.
\end{abstract}

\maketitle

\newpage 

Detecting novel magnetic interactions in spintronic materials is crucial for advancing our understanding of magnetization dynamics~\cite{Tserkovnyak2018, DasSarma04, Gambardell09, Jungwirth2018, Parvini}. The van der Waals ferromagnet Fe$_3$GeTe$_2$ is anticipated to host significant four-spin chiral interactions that may underlie the formation of non-collinear magnetic textures~\cite{Ado2020, Ado2021, Rakhmanova2022, Ado2022, Zheng2023, Heinze2011}.

\textit{Ab initio} studies of Fe$_3$GeTe$_2$ suggest that its magnon excitations can be effectively described by a ferrimagnet model, featuring a small imbalance in the average spin polarization between two magnetic sublattices~\cite{Reinhard06, Rudenko}. This leads to a magnon gap of approximately 100~meV~\cite{Akhiezer}, providing a possibility to explore unconventional magnon dynamics.

In this Letter, we propose that the decay of gapped magnons ($\beta$-magnons) in Fe$_3$GeTe$_2$ is predominantly governed by the four-spin chiral interaction. A distinctive signature of this decay channel is the cascade-like generation of magnons with large wave vectors. We further suggest that this process provides a direct and experimentally accessible route for detecting four-spin chiral interactions in Fe$_3$GeTe$_2$.

While numerous experimental reports have identified non-collinear magnetic textures in Fe$_3$GeTe$_2$~\cite{Ding2020, Meijer2020, Park2021}, the microscopic origin of these textures remains under debate. The D$_\textrm{3h}$ point group symmetry of Fe$_3$GeTe$_2$ forbids all Lifshitz invariants in the micromagnetic energy, making it unlikely that Dzyaloshinskii--Moriya interactions (DMI) stabilize the observed textures~\cite{Manchon2020, Everschor-Sitte2019}. Instead, the four-spin chiral interaction, a higher-order term linear in magnetization gradients but quartic in spin operators~\cite{Ado2020, Lifschitz1941}, emerges as a compelling candidate. This interaction can arise from indirect exchange mediated by conduction electrons and can stabilize non-collinear magnetic states independently of DMI.

To date, no experimental approach has been proposed to unambiguously distinguish whether DMI or four-spin chiral interactions are responsible for the non-collinear ground states. Experimental detection of the four-spin chiral interaction requires techniques capable of resolving magnon dynamics with both high energy and momentum resolution. Resonant Inelastic X-ray Scattering (RIXS), especially in its time-resolved implementation, offers a powerful method for this purpose~\cite{Mitrano2024, RIXS, TimeRIXS, TimeRIXS2}. Specifically, we predict that gapped $\beta$-magnons with zero wave vector decay into three low-energy magnons with finite momenta, a process that cannot be driven by conventional exchange interactions but is uniquely enabled by the four-spin chiral interaction. This mechanism can be viewed as a nonlinear analog of the Dzyaloshinskii--Moriya interaction~\cite{Dzyaloshinsky1958, Moriya1960, Kuepferling2023, Liang2020, Mazurenko2021}.

We further propose that this relaxation process can be triggered optically and detected using pump-probe techniques~\cite{Reznik2020, Dabrowski2022, Vahaplar2009, Radu2011}. By monitoring the decay of the resonant Kittel mode into finite-momentum magnons, one can access the magnon dynamics characteristic of the four-spin interaction. To describe this relaxation channel, we adopt a kinetic approach based on the time evolution of the magnon density~\cite{Keldysh, Rammer, Kamenev, Lifshitz, Calzetta}, a framework widely used to study ultrafast dynamics in magnonics~\cite{Beaurepaire1996, Shah1999, Qiu2021, Hennecke, Pogrebna, FeO, Bigot2005, RasingRMP, Rasing2014, Rasing2005, Atxitia, Mendil, Mentink2012, Mentink2015, Itin2015, Dolgikh23, Davies20}.

Our results identify the four-spin chiral interaction as the key driver of a novel magnon relaxation channel. If confirmed experimentally, this would provide not only direct evidence of four-spin interactions but also open new avenues for controlling complex magnetic dynamics in low-dimensional materials~\cite{Burch2018, Gibertini2019, Geim2013}. 

\begin{figure}[ht]
  \centering
  \subfigure[]{
    \includegraphics[width=0.6\columnwidth]{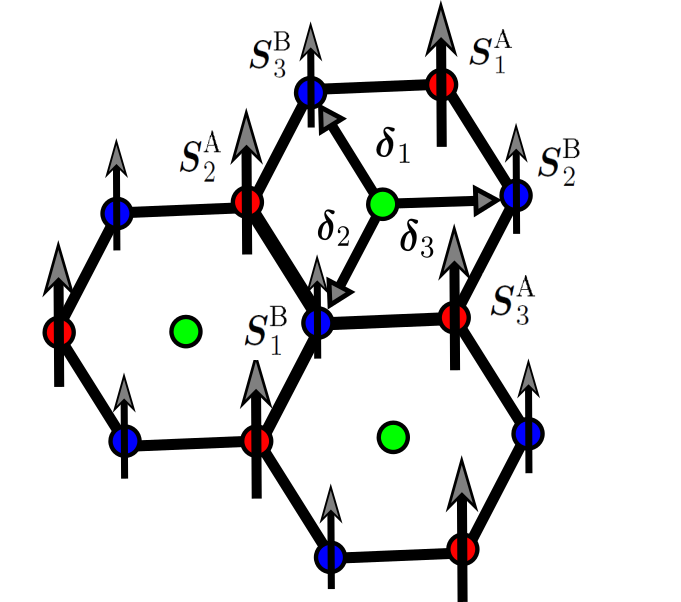}
    \label{fig:Lattice}
  }
  \\
  \subfigure[]{
    \includegraphics[width=0.75\columnwidth]{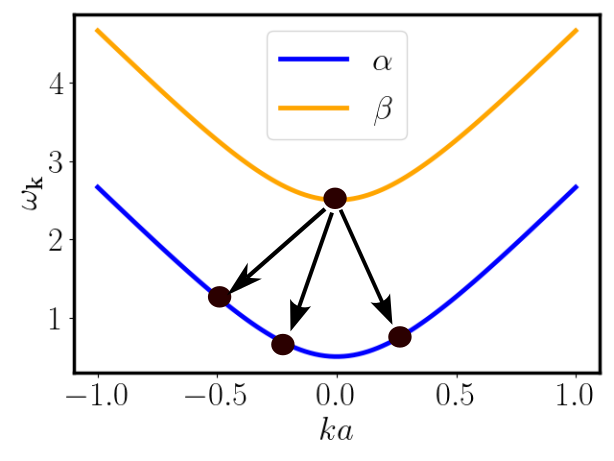}
    \label{fig:spectrumferrihoney2d}
  }
  \caption{(a) An effective honeycomb spin lattice for a ferrimagnet with different spins on sub-lattices A (blue) $S^{\textrm{A}}$ and B (red) $S^{\textrm{B}}$. Green non-magnetic atoms in the center of hexagons form a triangular dual lattice that is used to define the 4-spin chiral interaction on a lattice. (b) Characteristic magnon spectra of a ferrimagnet model. The gap between $\alpha$ and $\beta$ magnons is set by $J(S^{\textrm{A}}-S^{\textrm{B}})$, where $J$ is a nearest neighbour exchange energy.}
\end{figure}

\textit{Model}. --- We employ a classical effective Heisenberg model for a ferrimagnet on a two-dimensional honeycomb lattice shown in Fig.~\ref{fig:Lattice}a,
\begin{equation}
{\cal H}= -J\sum_{\langle \textit{i j} \rangle}\bs{S}_{\textit{i}}\cdot\bs{S}_{\textit{j}} -H \sum_{\textit{i}}S^{z}_{\textit{i}}  -h(t)\sum_{\textit{i}}S_{i}^{x}+{\cal H}_{\textrm{4S}},
\label{eq:HamSpin}
\end{equation}
where the spins on A and B sub-lattices have slightly different effective values $S^{\textrm{A}}$ and $S^{\textrm{B}}$, ($S^{\textrm{A}}>S^{\textrm{B}}$). The main interaction in the model is the ferromagnetic exchange interaction $J>0$, while the term ${\cal H}_{\textrm{4S}}$ stands for a weak 4-spin chiral interaction that we specify below. Constant external field $H$ is directed perpendicular to the plane. Weak time dependent in-plane field $h(t)$ ($|h(t)|\ll J,H$) is induced by electromagnetic waves to excite magnons. For a sake of simplicity we ignore anisotropy terms that play a role that is similar to that of external field.  

The sum $\langle ij \rangle$ in Eq.~(\ref{eq:HamSpin}) extends over the nearest neighbours. The positions of three nearest neighbors to an $A$ atom are obtained by the translations along the vectors $\bb{\delta}_\alpha$. For a sake of definiteness we choose  
$\bb{\delta}_{1} = a\begin{pmatrix}1, & 0\end{pmatrix}$, $\bb{\delta}_{2} = \frac{a}{2}\begin{pmatrix}-1, & \sqrt{3}\end{pmatrix}$, and $\bb{\delta}_{3} = \frac{a}{2}\begin{pmatrix}-1, & -\sqrt{3}\end{pmatrix}\ $.

The 4-spin chiral interaction ${\cal H}_{\textrm{4S}}$ on the lattice is conveniently defined as the placket summation,
\begin{align}
{\cal H}_{\textrm{4S}}=&\frac{A_{\textrm{D}}}{a^4}\sum_{\bs{R}\in O}\lt[\frac{1}{(S^{\textrm{A}})^{3}S^{\textrm{B}}}\prod_{\alpha=1}^{3}\bb{\delta}_{\alpha}\cdot\bs{S}_{\bs{R}+\bb{\delta}_{\alpha}}^{\textrm{A}}\sum_{\beta}\bb{\delta}_{\beta}\cdot\bs{S}_{\bs{R}+\bb{\delta}_{\beta}}^{\textrm{B}}\rt. \n\\
&-\lt.\frac{1}{S^{\textrm{A}}(S^{\textrm{B}})^{3}}\prod_{\beta=1}^{3}\bb{\delta}_{\beta}\cdot\bs{S}_{\bs{R}+\bb{\delta}_{\beta}}^{\textrm{B}}\sum_{\alpha}\bs{\delta}_{\alpha}\cdot\bs{S}_{\bs{R}+\bs{\delta}_{\alpha}}^{\textrm{A}}\rt],
\label{eq:Ham4S}
\end{align}
where $\bs{R} \in O$ are referring to the coordinates of the dual lattice (hexagon centers) while $A_{\textrm{D}}$ is a characteristic energy of the interaction strength. 

The free energy function of the micromagnetic description can be defined in terms of the smooth unit vector field $\bs{n}(\bs{r})$ that equals $\bs{S}^A_\bs{r}/S^A$ or $\bs{S}^B_\bs{r}/S^B$ on the corresponding sub-lattices. In terms of the vector field one can represent the interaction energy as
\be
\label{H4Sn}
{\cal H}_{{\textrm{4S}}}= A_D\int d^3\bs{r}\, n_1 n_2 n_3 \dv n, \qquad n_\alpha= \bb{\delta}_\alpha\cdot \bb{n},
\e
which cannot be reduced to the integrals from Lifshitz invariants. It can be formally shown that the latter are forbidden for systems with $D_{\textrm{3h}}$ point group symmetry \cite{Ado2020, Ado2021}. The term of Eq.~(\ref{H4Sn}) reflects the absence of crystal inversion symmetry and can become responsible for the stability of non-collinear magnetic textures: spin spirals or bi-merons \cite{Ado2021, Ado2022}. 

We assume collinear ground state of the system with all spins pointing out in $z$ directions by the external field $H$. In this case, the linearized magnon Hamiltonian reads
\begin{align}
{\cal H}_{0}=&\sum_{\bs{k}}\lt[\omega^{\alpha}_\bs{k}\alpha_{\bs{k}}\h\alpha_{\bs{k}}\0+\omega^{\beta}_{\bs{k}}\beta_{\bs{k}}\h\beta_{\bs{k}}\0\rt]\n\\
&-h(t)\lt[A(\alpha_{0}\0 +\alpha_{0}\h)+B(\beta_{0}\0+\beta_{0}\h)\rt],
\end{align}
where $A$ and $B$ are the exchange coefficients provided in Ref.~\onlinecite{Supp}. 

The corresponding magnon spectra, which are defined by the Hamiltonian ${\cal H}_{0}$ with $h(t)=0$, are represented by two branches that are formally described by the second quantization operators $\alpha_{\bs{k}}$ and $\beta_{\bs{k}}$ in $\bs{k}$ space \cite{Supp}. The spectra consist of two quadratic branches $\omega^{\alpha}_{\bs{k}}, \omega^{\beta}_{\bs{k}}$ shown in Fig. \ref{fig:spectrumferrihoney2d}. The spectral gap is given by $\Delta=\omega^\beta_{0}-\omega^\alpha_{0}= J(S^{\textrm{A}}-S^{\textrm{B}})$, where $\omega^{\alpha,\beta}_0=\omega^{\alpha,\beta}_{\bs{k}=0}$. Note that in the absence of anisotropy, the gap of $\alpha$ magnons, $\omega^\alpha_{0}$, is set solely by the external field $H$. 

In order to describe magnetization dynamics we introduce magnon distribution functions $N^{\alpha}_{\bs{k}}=\la \alpha_{\bs{k}}\h (t)\alpha_{\bs{k}}\0 (t)\ra$ and $N^{\beta}_{\bs{k}}=\la \beta_{\bs{k}}\h (t)\beta_{\bs{k}}\0 (t)\ra$. The time dependent problem described by ${\cal H}_{0}$ can be solved exactly by considering the Heisenberg equations of motion for $\alpha_0=\alpha_{\bs{k}=0}$ and $\beta_0=\beta_{\bs{k}=0}$,
\beml
\label{Heis}
\begin{align}
d\alpha_{0}/dt &= -i(\omega_{0}^{\alpha}\alpha_{0}-h(t)A), \\
d\beta_{0}/dt &= -i(\omega_{0}^{\beta}\beta_{0}-h(t)B),
\end{align}
\eml
while the kinetic equations can be cast in the following form
\beml
\begin{align}
\label{kinalphaforkzero}
dN_{\bs{k}}^{\alpha}/dt-\delta_{\bs{k},0} F^{\alpha}(t)= &I^{{\textrm{ex}}}_{\alpha\alpha} + I^{{\textrm{ex}}}_{\alpha\beta} + I^{{\textrm{4S}}}_{\alpha},\\
dN_{\bs{k}}^{\beta}/dt-\delta_{\bs{k},0} F^{\beta}(t)= &I^{{\textrm{ex}}}_{\beta\beta} + I^{{\textrm{ex}}}_{\beta\alpha} + I^{{\textrm{4S}}}_{\beta},
\label{kinbetaforkzero}  
\end{align}
\eml
where we have introduced the generalized forces 
\beml
\begin{align}
\label{eq:Falpha}
F^{\alpha}(t) = 2A^2 h(t)\int_{0}^{t}h(t')\cos(\omega^{\alpha}_{\bs{k}=0}(t-t'))dt',\\
F^{\beta}(t) = 2B^2 h(t)\int_{0}^{t}h(t')\cos(\omega^{\beta}_{\bs{k}=0}(t-t'))dt',
\label{eq:Fbeta}
\end{align}
\eml
which follow from the solution of Eqs.~(\ref{Heis}). 

The collision integrals $I^{{\textrm{ex}}}$ and $I^{{\textrm{4S}}}$ describe exchange and 4-spin chiral interactions, correspondingly \cite{Supp}. Both interactions lead to the redistribution of energy and momentum among magnons with, however, principally different constraints.  

The Heisenberg exchange interaction, represented by $I^\textrm{ex}$, conserves the total number of magnons. As the result the exchange interaction does only contain magnon scattering processes and cannot describe magnon decay, which intrinsically require relativistic processes. Specifically we are looking for a decay of $\beta$ magnon with $\bs{k}=0$ into $\alpha$ magnons with finite wave vectors. It can be shown that usual 2-spin relativistic interaction such as DMI do also forbid such processes. One needs 4-spin chiral interaction for such a process to occur. This can be seen directly from the structure of the collision integral, $I^{{\textrm{4S}}}$, presented int the Supplemental Material. 

\begin{figure}[!htb]
 \centering
 \begin{tikzpicture}
   \node[inner sep=0pt] (main) at (0,0) {\includegraphics[width=0.48\textwidth]{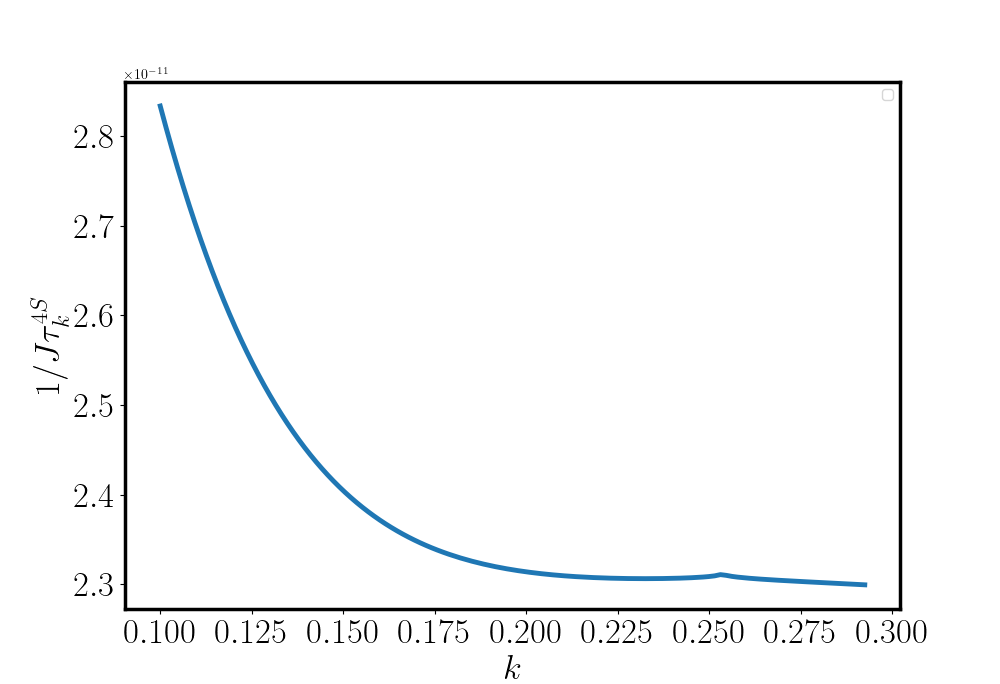}};
   \node[inner sep=0pt] at ([xshift=5.5cm, yshift=-2.5cm] main.north west) {\includegraphics[width=0.25\textwidth]{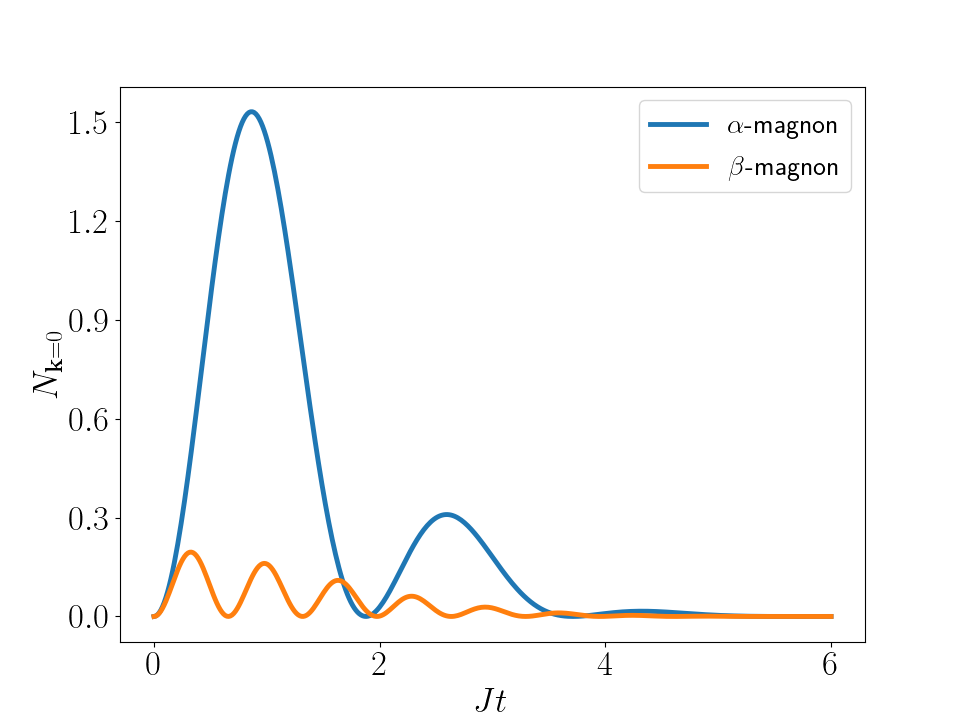}};
 \end{tikzpicture}
 \caption{Scattering time of 4-spin interaction for different wave vectors. Inset: dynamics of the number of magnons with $\mathbf{k}=0$ of sort $\alpha$ with a Gaussian pulse with $Jt_{p}=3$, $h_{0}/J=0.1$, $H/J = 0.5$.}
 \label{fig:scatteringtime}
\end{figure}

The magnetization dynamics in van der Waals magnets is normally investigated in pump-probe experiments. In these experiments, the magnons are periodically excited with laser pump pulses, while the resulting magnetization is optically probed with the same periodicity to exclude noise. Each pump pulse can can be regarded as Gaussian $h(t) = h_0 \exp\left(-\frac{t^2}{t_\textrm{p}^2}\right)$, where $h_0$ is the amplitude and $t_\textrm{p}$ is the duration of the pulse. The pulses pump both $\alpha$ and $\beta$ magnons in the Kittel mode, $\bs{k}=0$. The dynamics of the Kittel mode of $\alpha$ and $\beta$ magnons under the influence of the Gaussian pulse (obtained from Eqs.~(\ref{Heis})) is shown in the inset of Fig. \ref{fig:scatteringtime}. The excited magnons reach equilibrium via the scattering and decay processes. The scattering processes are responsible for the oscillations of magnon density on timescales on the order of $1/J$. 

The decay  of $\beta$ magnons with  $\bs{k}=0$ to $\alpha$ magnons with finite $\bs{k}$ values does happen on much longer time scales $1/A_\textrm{D}$ that correspond to the 4-spin chiral interaction as shown in Fig. \ref{fig:spectrumferrihoney2d}. This decay is governed by the collision term  $I^{{\textrm{4S}}}_{\alpha}$. The energy conservation for this decay process reads $\Delta = \omega_{\bs{k}_{1}}^{\alpha} +\omega_{\bs{k}_{2}}^{\alpha} + \omega_{\bs{k}_{3}}^{\alpha}$, subject to the constraint $\mathbf{k}_{1} + \mathbf{k}_{2} + \mathbf{k}_{3} = 0$. The Hamiltonian incorporating these terms has the following form:
\begin{equation}
\begin{aligned}
&{\cal H}_{{\textrm{4S}}} = \frac{A_{\textrm{D}}}{32a^{4}(S^{\textrm{A}}S^{\textrm{B}})^{3/2}} \Bigg[\sum_{\bs{k}_{1}\bs{k}_{2}\bs{k}_{3}\bs{k}_{4}}\delta(\bs{k}_{1}+\bs{k}_{2}+\bs{k}_{3}+\bs{k}_{4}) \\
&\tilde{U}_{\bs{k}_{1},\bs{k}_{2},\bs{k}_{3},\bs{k}_{4}}\alpha_{\bs{k}_{1}}^{\dagger}\alpha_{\bs{k}_{2}}^{\dagger}\alpha_{\bs{k}_{3}}^{\dagger}\beta_{\bs{k}_{4}}\Bigg] +h.c., \\
\end{aligned}
\end{equation}
where $\tilde{U}_{1}$ is explicitly presented in the Supplementary Material. Other terms in the 4-spin Hamiltonian with different structures are omitted due to the condition $A_{\textrm{D}} \ll J$.

\begin{figure}[!htb]
   \begin{minipage}{0.44\textwidth}
     \centering
     \includegraphics[width=1\linewidth]{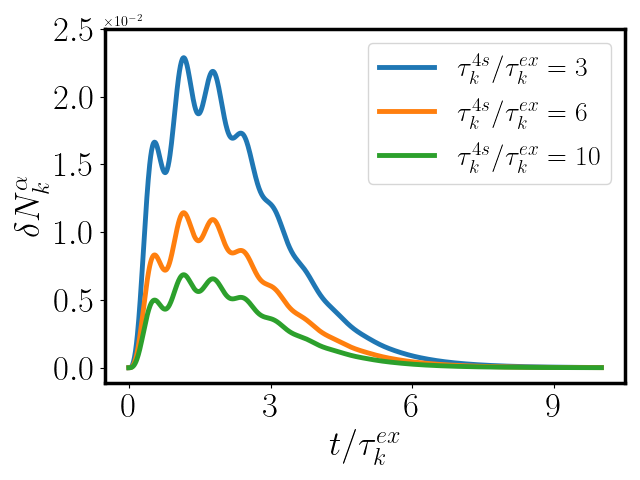}
     \caption{Number's deviation of $\alpha$-magnons from equilibrium with $\bs{k}\neq 0$ for different scattering time $\tau^{{\textrm{4S}}}$ according the formula (\ref{eq:FinalKin}).}
     \label{fig:solkinalpha}
   \end{minipage}\hfill
\end{figure}

In order to solve Eq~(\ref{kinalphaforkzero}) we consider deviation from Bose equilibrium $N^{\alpha}_{\bs{k}}=N^{eq,\alpha}_{\bs{k}}+\delta N^{\alpha}_{\bs{k}}$,
and  $N^{\beta}_{\bs{k}}=N^{eq,\beta}_{\bs{k}}+N_{\bs{k}=0}^{\beta}(t)$, where $N_{\bs{k}=0}^{\beta}(t)$ is a solution of Eq~(\ref{kinbetaforkzero})  with  a Gaussian pulse and depicted in Fig. (\ref{fig:scatteringtime}).
Taking into account exchange interaction is larger than 4-spin interaction, we obtain an equation for $\delta N^{\alpha}$ 
\begin{equation}
\frac{d\delta N_{\bs{k}}^{\alpha}}{dt}=-\frac{\delta N_{\bs{k}}^{\alpha}}{\tau_{\bs{k}}^{{\textrm{ex}}}}+\frac{N_{\bs{k}=0}^{\beta}(t)}{\tau_{\bs{k}}^{\textrm{4S}}}.
\label{eq:FinalKin}
\end{equation}
This kinetic equation can be interpreted in the following way: the 4-spin interaction provides a decay channel that excites low energy magnons with finite wave vectors. These magnons are then equilibrated to a time dependent thermal distrubution by means of the Heisenberg exchange. 

The scattering time of the 4-spin interaction $\tau_{\bs{k}}^{\textrm{4S}}$ shown in Fig. (\ref{fig:scatteringtime}), we can see that this process has a non-zero channel. On the Fig. (\ref{fig:solkinalpha}) a solution for $\delta N_{\bs{k}}^{\alpha}$ that shows a peak in magnon excitation occurs at times on the order of $\tau^{\textrm{ex}}_{\bs{k}}$ . 

The lifetime of a $\beta$-magnon in this process is proportional to the square of the energy of the 4-spin interaction divided by the exchange energy, given by

\begin{equation}
\frac{1}{\tau_{\beta}}=C\frac{A_{D}^{2}}{J},
\label{eq:lifetimebetamagnon}
\end{equation}
where $C$ is a dimensionless coefficient provided in the Supplementary Material.

To quantify the $\beta$-magnon lifetime, we evaluate Eq.~(\ref{eq:lifetimebetamagnon}) using parameters relevant to Fe$_3$GeTe$_2$. For a honeycomb ferrimagnet with effective spins $S^{\mathrm{A}} = 2.2$ and $S^{\mathrm{B}} = 1.2$, and an exchange energy $J \approx 100\,\mathrm{meV}$, the prefactor in Eq.~(\ref{eq:lifetimebetamagnon}) yields a dimensionless constant $C \approx 1.988 \times 10^{-7}$. Assuming a four-spin chiral interaction strength of $A_D = 0.01 J$, the relaxation rate is $\tau_{\beta}^{-1} \approx 480.5\,\mathrm{s}^{-1}$, corresponding to a lifetime $\tau_{\beta} \approx 2.081\,\mathrm{ms}$. For a weaker interaction, $A_D = 0.001 J$, the lifetime extends to $\tau_{\beta} \approx 0.2081\,\mathrm{s}$. Such millisecond timescales are readily accessible with time-resolved resonant inelastic X-ray scattering (TR-RIXS)~\cite{Mitrano2024, RIXS, TimeRIXS, TimeRIXS2}, which offers femtosecond to nanosecond resolution, or with optical pump-probe techniques. This slow relaxation, driven by the weak four-spin chiral interaction, provides a clear and distinctive experimental signature for testing the proposed mechanism.

\textit{Summary}.---We have proposed an all-optical approach to detect four-spin chiral interactions in two-dimensional honeycomb ferrimagnets with D$_{3h}$ magnetic lattice symmetry, focusing on an effective model of the van der Waals ferromagnet Fe$_3$GeTe$_2$. Unlike conventional inversion symmetry breaking described by Lifshitz invariants, the symmetry of this material necessitates a more complex four-spin chiral interaction that cannot be reduced to the Dzyaloshinskii--Moriya interaction.

Through a theoretical analysis of magnon dynamics, we have demonstrated that the four-spin chiral interaction uniquely enables the decay of gapped $\beta$-magnons with zero wave vector into three low-energy $\alpha$-magnons with finite wave vectors. We propose that this distinctive relaxation channel can be detected using optical pump-probe experiments. Notably, this decay process cannot be induced by Heisenberg exchange, magnetic anisotropy, or Dzyaloshinskii--Moriya interactions. Our findings provide new insights into the complex magnetic dynamics of van der Waals magnetic materials and open promising directions for experimental exploration.

\textit{Acknowledgments}.---We thank Ivan Ado for valuable discussions. This work was supported by the European Union’s Horizon 2020 research and innovation program under the Marie Skłodowska-Curie grant agreement No.~873028 and by the Netherlands Organisation for Scientific Research (NWO) through the Spinoza Prize by MIK.
	
\bibliographystyle{apsrev4-1}
\bibliography{Bibliography}

\begin{thebibliography}{60}%
\makeatletter
\providecommand \@ifxundefined [1]{%
 \@ifx{#1\undefined}
}%
\providecommand \@ifnum [1]{%
 \ifnum #1\expandafter \@firstoftwo
 \else \expandafter \@secondoftwo
 \fi
}%
\providecommand \@ifx [1]{%
 \ifx #1\expandafter \@firstoftwo
 \else \expandafter \@secondoftwo
 \fi
}%
\providecommand \natexlab [1]{#1}%
\providecommand \enquote  [1]{``#1''}%
\providecommand \bibnamefont  [1]{#1}%
\providecommand \bibfnamefont [1]{#1}%
\providecommand \citenamefont [1]{#1}%
\providecommand \href@noop [0]{\@secondoftwo}%
\providecommand \href [0]{\begingroup \@sanitize@url \@href}%
\providecommand \@href[1]{\@@startlink{#1}\@@href}%
\providecommand \@@href[1]{\endgroup#1\@@endlink}%
\providecommand \@sanitize@url [0]{\catcode `\\12\catcode `\$12\catcode `\&12\catcode `\#12\catcode `\^12\catcode `\_12\catcode `\%12\relax}%
\providecommand \@@startlink[1]{}%
\providecommand \@@endlink[0]{}%
\providecommand \url  [0]{\begingroup\@sanitize@url \@url }%
\providecommand \@url [1]{\endgroup\@href {#1}{\urlprefix }}%
\providecommand \urlprefix  [0]{URL }%
\providecommand \Eprint [0]{\href }%
\providecommand \doibase [0]{http://dx.doi.org/}%
\providecommand \selectlanguage [0]{\@gobble}%
\providecommand \bibinfo  [0]{\@secondoftwo}%
\providecommand \bibfield  [0]{\@secondoftwo}%
\providecommand \translation [1]{[#1]}%
\providecommand \BibitemOpen [0]{}%
\providecommand \bibitemStop [0]{}%
\providecommand \bibitemNoStop [0]{.\EOS\space}%
\providecommand \EOS [0]{\spacefactor3000\relax}%
\providecommand \BibitemShut  [1]{\csname bibitem#1\endcsname}%
\let\auto@bib@innerbib\@empty
\bibitem [{\citenamefont {Baltz}\ \emph {et~al.}(2018)\citenamefont {Baltz}, \citenamefont {Manchon}, \citenamefont {Tsoi}, \citenamefont {Moriyama}, \citenamefont {Ono},\ and\ \citenamefont {Tserkovnyak}}]{Tserkovnyak2018}%
  \BibitemOpen
  \bibfield  {author} {\bibinfo {author} {\bibfnamefont {V.}~\bibnamefont {Baltz}}, \bibinfo {author} {\bibfnamefont {A.}~\bibnamefont {Manchon}}, \bibinfo {author} {\bibfnamefont {M.}~\bibnamefont {Tsoi}}, \bibinfo {author} {\bibfnamefont {T.}~\bibnamefont {Moriyama}}, \bibinfo {author} {\bibfnamefont {T.}~\bibnamefont {Ono}}, \ and\ \bibinfo {author} {\bibfnamefont {Y.}~\bibnamefont {Tserkovnyak}},\ }\href {\doibase 10.1103/RevModPhys.90.015005} {\bibfield  {journal} {\bibinfo  {journal} {Rev. Mod. Phys.}\ }\textbf {\bibinfo {volume} {90}},\ \bibinfo {pages} {015005} (\bibinfo {year} {2018})}\BibitemShut {NoStop}%
\bibitem [{\citenamefont {\ifmmode \check{Z}\else \v{Z}\fi{}uti\ifmmode~\acute{c}\else \'{c}\fi{}}\ \emph {et~al.}(2004)\citenamefont {\ifmmode \check{Z}\else \v{Z}\fi{}uti\ifmmode~\acute{c}\else \'{c}\fi{}}, \citenamefont {Fabian},\ and\ \citenamefont {Das~Sarma}}]{DasSarma04}%
  \BibitemOpen
  \bibfield  {author} {\bibinfo {author} {\bibfnamefont {I.}~\bibnamefont {\ifmmode \check{Z}\else \v{Z}\fi{}uti\ifmmode~\acute{c}\else \'{c}\fi{}}}, \bibinfo {author} {\bibfnamefont {J.}~\bibnamefont {Fabian}}, \ and\ \bibinfo {author} {\bibfnamefont {S.}~\bibnamefont {Das~Sarma}},\ }\href {\doibase 10.1103/RevModPhys.76.323} {\bibfield  {journal} {\bibinfo  {journal} {Rev. Mod. Phys.}\ }\textbf {\bibinfo {volume} {76}},\ \bibinfo {pages} {323} (\bibinfo {year} {2004})}\BibitemShut {NoStop}%
\bibitem [{\citenamefont {Manchon}\ \emph {et~al.}(2019)\citenamefont {Manchon}, \citenamefont {\ifmmode~\check{Z}\else \v{Z}\fi{}elezn\'y}, \citenamefont {Miron}, \citenamefont {Jungwirth}, \citenamefont {Sinova}, \citenamefont {Thiaville}, \citenamefont {Garello},\ and\ \citenamefont {Gambardella}}]{Gambardell09}%
  \BibitemOpen
  \bibfield  {author} {\bibinfo {author} {\bibfnamefont {A.}~\bibnamefont {Manchon}}, \bibinfo {author} {\bibfnamefont {J.}~\bibnamefont {\ifmmode~\check{Z}\else \v{Z}\fi{}elezn\'y}}, \bibinfo {author} {\bibfnamefont {I.~M.}\ \bibnamefont {Miron}}, \bibinfo {author} {\bibfnamefont {T.}~\bibnamefont {Jungwirth}}, \bibinfo {author} {\bibfnamefont {J.}~\bibnamefont {Sinova}}, \bibinfo {author} {\bibfnamefont {A.}~\bibnamefont {Thiaville}}, \bibinfo {author} {\bibfnamefont {K.}~\bibnamefont {Garello}}, \ and\ \bibinfo {author} {\bibfnamefont {P.}~\bibnamefont {Gambardella}},\ }\href {\doibase 10.1103/RevModPhys.91.035004} {\bibfield  {journal} {\bibinfo  {journal} {Rev. Mod. Phys.}\ }\textbf {\bibinfo {volume} {91}},\ \bibinfo {pages} {035004} (\bibinfo {year} {2019})}\BibitemShut {NoStop}%
\bibitem [{\citenamefont {Jungwirth}\ \emph {et~al.}(2018)\citenamefont {Jungwirth}, \citenamefont {Sinova}, \citenamefont {Manchon}, \citenamefont {Marti}, \citenamefont {Wunderlich},\ and\ \citenamefont {Felser}}]{Jungwirth2018}%
  \BibitemOpen
  \bibfield  {author} {\bibinfo {author} {\bibfnamefont {T.}~\bibnamefont {Jungwirth}}, \bibinfo {author} {\bibfnamefont {J.}~\bibnamefont {Sinova}}, \bibinfo {author} {\bibfnamefont {A.}~\bibnamefont {Manchon}}, \bibinfo {author} {\bibfnamefont {X.}~\bibnamefont {Marti}}, \bibinfo {author} {\bibfnamefont {J.}~\bibnamefont {Wunderlich}}, \ and\ \bibinfo {author} {\bibfnamefont {C.}~\bibnamefont {Felser}},\ }\href {\doibase 10.1038/s41567-018-0063-6} {\bibfield  {journal} {\bibinfo  {journal} {Nature Physics}\ }\textbf {\bibinfo {volume} {14}},\ \bibinfo {pages} {200} (\bibinfo {year} {2018})}\BibitemShut {NoStop}%
\bibitem [{\citenamefont {Parvini}\ \emph {et~al.}(2020)\citenamefont {Parvini}, \citenamefont {Bittencourt},\ and\ \citenamefont {Kusminskiy}}]{Parvini}%
  \BibitemOpen
  \bibfield  {author} {\bibinfo {author} {\bibfnamefont {T.~S.}\ \bibnamefont {Parvini}}, \bibinfo {author} {\bibfnamefont {V.~A. S.~V.}\ \bibnamefont {Bittencourt}}, \ and\ \bibinfo {author} {\bibfnamefont {S.~V.}\ \bibnamefont {Kusminskiy}},\ }\href {\doibase 10.1103/PhysRevResearch.2.022027} {\bibfield  {journal} {\bibinfo  {journal} {Phys. Rev. Research}\ }\textbf {\bibinfo {volume} {2}},\ \bibinfo {pages} {022027} (\bibinfo {year} {2020})}\BibitemShut {NoStop}%
\bibitem [{\citenamefont {Ado}\ \emph {et~al.}(2020)\citenamefont {Ado}, \citenamefont {Qaiumzadeh}, \citenamefont {Brataas},\ and\ \citenamefont {Titov}}]{Ado2020}%
  \BibitemOpen
  \bibfield  {author} {\bibinfo {author} {\bibfnamefont {I.~A.}\ \bibnamefont {Ado}}, \bibinfo {author} {\bibfnamefont {A.}~\bibnamefont {Qaiumzadeh}}, \bibinfo {author} {\bibfnamefont {A.}~\bibnamefont {Brataas}}, \ and\ \bibinfo {author} {\bibfnamefont {M.}~\bibnamefont {Titov}},\ }\href {\doibase 10.1103/PhysRevB.101.161403} {\bibfield  {journal} {\bibinfo  {journal} {Phys. Rev. B}\ }\textbf {\bibinfo {volume} {101}},\ \bibinfo {pages} {161403} (\bibinfo {year} {2020})}\BibitemShut {NoStop}%
\bibitem [{\citenamefont {Ado}\ \emph {et~al.}(2021)\citenamefont {Ado}, \citenamefont {Tchernyshyov},\ and\ \citenamefont {Titov}}]{Ado2021}%
  \BibitemOpen
  \bibfield  {author} {\bibinfo {author} {\bibfnamefont {I.~A.}\ \bibnamefont {Ado}}, \bibinfo {author} {\bibfnamefont {O.}~\bibnamefont {Tchernyshyov}}, \ and\ \bibinfo {author} {\bibfnamefont {M.}~\bibnamefont {Titov}},\ }\href {\doibase 10.1103/PhysRevLett.127.127204} {\bibfield  {journal} {\bibinfo  {journal} {Phys. Rev. Lett.}\ }\textbf {\bibinfo {volume} {127}},\ \bibinfo {pages} {127204} (\bibinfo {year} {2021})}\BibitemShut {NoStop}%
\bibitem [{\citenamefont {Rakhmanova}\ \emph {et~al.}(2022)\citenamefont {Rakhmanova}, \citenamefont {Osipov}, \citenamefont {Ilyin}, \citenamefont {Shushakova}, \citenamefont {Ado}, \citenamefont {Iorsh},\ and\ \citenamefont {Titov}}]{Rakhmanova2022}%
  \BibitemOpen
  \bibfield  {author} {\bibinfo {author} {\bibfnamefont {G.}~\bibnamefont {Rakhmanova}}, \bibinfo {author} {\bibfnamefont {A.}~\bibnamefont {Osipov}}, \bibinfo {author} {\bibfnamefont {D.}~\bibnamefont {Ilyin}}, \bibinfo {author} {\bibfnamefont {I.}~\bibnamefont {Shushakova}}, \bibinfo {author} {\bibfnamefont {I.~A.}\ \bibnamefont {Ado}}, \bibinfo {author} {\bibfnamefont {I.}~\bibnamefont {Iorsh}}, \ and\ \bibinfo {author} {\bibfnamefont {M.}~\bibnamefont {Titov}},\ }\href {\doibase 10.1103/PhysRevB.105.L020401} {\bibfield  {journal} {\bibinfo  {journal} {Phys. Rev. B}\ }\textbf {\bibinfo {volume} {105}},\ \bibinfo {pages} {L020401} (\bibinfo {year} {2022})}\BibitemShut {NoStop}%
\bibitem [{\citenamefont {Ado}\ \emph {et~al.}(2022)\citenamefont {Ado}, \citenamefont {Rakhmanova}, \citenamefont {Zezyulin}, \citenamefont {Iorsh},\ and\ \citenamefont {Titov}}]{Ado2022}%
  \BibitemOpen
  \bibfield  {author} {\bibinfo {author} {\bibfnamefont {I.~A.}\ \bibnamefont {Ado}}, \bibinfo {author} {\bibfnamefont {G.}~\bibnamefont {Rakhmanova}}, \bibinfo {author} {\bibfnamefont {D.~A.}\ \bibnamefont {Zezyulin}}, \bibinfo {author} {\bibfnamefont {I.}~\bibnamefont {Iorsh}}, \ and\ \bibinfo {author} {\bibfnamefont {M.}~\bibnamefont {Titov}},\ }\href {\doibase 10.1103/PhysRevB.106.144407} {\bibfield  {journal} {\bibinfo  {journal} {Phys. Rev. B}\ }\textbf {\bibinfo {volume} {106}},\ \bibinfo {pages} {144407} (\bibinfo {year} {2022})}\BibitemShut {NoStop}%
\bibitem [{\citenamefont {Zheng}\ \emph {et~al.}(2023)\citenamefont {Zheng}, \citenamefont {Kiselev}, \citenamefont {Rybakov}, \citenamefont {Yang}, \citenamefont {Shi}, \citenamefont {Bl{\"u}gel},\ and\ \citenamefont {Dunin-Borkowski}}]{Zheng2023}%
  \BibitemOpen
  \bibfield  {author} {\bibinfo {author} {\bibfnamefont {F.}~\bibnamefont {Zheng}}, \bibinfo {author} {\bibfnamefont {N.~S.}\ \bibnamefont {Kiselev}}, \bibinfo {author} {\bibfnamefont {F.~N.}\ \bibnamefont {Rybakov}}, \bibinfo {author} {\bibfnamefont {L.}~\bibnamefont {Yang}}, \bibinfo {author} {\bibfnamefont {W.}~\bibnamefont {Shi}}, \bibinfo {author} {\bibfnamefont {S.}~\bibnamefont {Bl{\"u}gel}}, \ and\ \bibinfo {author} {\bibfnamefont {R.~E.}\ \bibnamefont {Dunin-Borkowski}},\ }\href@noop {} {\bibfield  {journal} {\bibinfo  {journal} {Nature}\ }\textbf {\bibinfo {volume} {623}},\ \bibinfo {pages} {718} (\bibinfo {year} {2023})}\BibitemShut {NoStop}%
\bibitem [{\citenamefont {Heinze}\ \emph {et~al.}(2011)\citenamefont {Heinze}, \citenamefont {von Bergmann}, \citenamefont {Menzel}, \citenamefont {Brede}, \citenamefont {Kubetzka}, \citenamefont {Wiesendanger}, \citenamefont {Bihlmayer},\ and\ \citenamefont {Bl{\"u}gel}}]{Heinze2011}%
  \BibitemOpen
  \bibfield  {author} {\bibinfo {author} {\bibfnamefont {S.}~\bibnamefont {Heinze}}, \bibinfo {author} {\bibfnamefont {K.}~\bibnamefont {von Bergmann}}, \bibinfo {author} {\bibfnamefont {M.}~\bibnamefont {Menzel}}, \bibinfo {author} {\bibfnamefont {J.}~\bibnamefont {Brede}}, \bibinfo {author} {\bibfnamefont {A.}~\bibnamefont {Kubetzka}}, \bibinfo {author} {\bibfnamefont {R.}~\bibnamefont {Wiesendanger}}, \bibinfo {author} {\bibfnamefont {G.}~\bibnamefont {Bihlmayer}}, \ and\ \bibinfo {author} {\bibfnamefont {S.}~\bibnamefont {Bl{\"u}gel}},\ }\href@noop {} {\bibfield  {journal} {\bibinfo  {journal} {Nature Physics}\ }\textbf {\bibinfo {volume} {7}},\ \bibinfo {pages} {713} (\bibinfo {year} {2011})}\BibitemShut {NoStop}%
\bibitem [{\citenamefont {Deiseroth}\ \emph {et~al.}(2006)\citenamefont {Deiseroth}, \citenamefont {Aleksandrov}, \citenamefont {Reiner}, \citenamefont {Kienle},\ and\ \citenamefont {Kremer}}]{Reinhard06}%
  \BibitemOpen
  \bibfield  {author} {\bibinfo {author} {\bibfnamefont {H.-J.}\ \bibnamefont {Deiseroth}}, \bibinfo {author} {\bibfnamefont {K.}~\bibnamefont {Aleksandrov}}, \bibinfo {author} {\bibfnamefont {C.}~\bibnamefont {Reiner}}, \bibinfo {author} {\bibfnamefont {L.}~\bibnamefont {Kienle}}, \ and\ \bibinfo {author} {\bibfnamefont {R.~K.}\ \bibnamefont {Kremer}},\ }\href {\doibase https://doi.org/10.1002/ejic.200501020} {\bibfield  {journal} {\bibinfo  {journal} {European Journal of Inorganic Chemistry}\ }\textbf {\bibinfo {volume} {2006}},\ \bibinfo {pages} {1561} (\bibinfo {year} {2006})},\ \Eprint {http://arxiv.org/abs/https://chemistry-europe.onlinelibrary.wiley.com/doi/pdf/10.1002/ejic.200501020} {https://chemistry-europe.onlinelibrary.wiley.com/doi/pdf/10.1002/ejic.200501020} \BibitemShut {NoStop}%
\bibitem [{\citenamefont {Pushkarev}\ \emph {et~al.}(2023)\citenamefont {Pushkarev}, \citenamefont {Badrtdinov}, \citenamefont {Iakovlev}, \citenamefont {Mazurenko},\ and\ \citenamefont {Rudenko}}]{Rudenko}%
  \BibitemOpen
  \bibfield  {author} {\bibinfo {author} {\bibfnamefont {G.~V.}\ \bibnamefont {Pushkarev}}, \bibinfo {author} {\bibfnamefont {D.~I.}\ \bibnamefont {Badrtdinov}}, \bibinfo {author} {\bibfnamefont {I.~A.}\ \bibnamefont {Iakovlev}}, \bibinfo {author} {\bibfnamefont {V.~V.}\ \bibnamefont {Mazurenko}}, \ and\ \bibinfo {author} {\bibfnamefont {A.~N.}\ \bibnamefont {Rudenko}},\ }\href {\doibase https://doi.org/10.1016/j.jmmm.2023.171456} {\bibfield  {journal} {\bibinfo  {journal} {Journal of Magnetism and Magnetic Materials}\ }\textbf {\bibinfo {volume} {588}},\ \bibinfo {pages} {171456} (\bibinfo {year} {2023})}\BibitemShut {NoStop}%
\bibitem [{\citenamefont {Akhiezer}\ \emph {et~al.}(1968)\citenamefont {Akhiezer}, \citenamefont {Baryakhtar},\ and\ \citenamefont {Peletminskii}}]{Akhiezer}%
  \BibitemOpen
  \bibfield  {author} {\bibinfo {author} {\bibfnamefont {A.~I.}\ \bibnamefont {Akhiezer}}, \bibinfo {author} {\bibfnamefont {V.~G.}\ \bibnamefont {Baryakhtar}}, \ and\ \bibinfo {author} {\bibfnamefont {S.~V.}\ \bibnamefont {Peletminskii}},\ }\href@noop {} {\emph {\bibinfo {title} {Spin waves}}}\ (\bibinfo  {publisher} {North-Holland Pub. Co.},\ \bibinfo {address} {Amsterdam},\ \bibinfo {year} {1968})\BibitemShut {NoStop}%
\bibitem [{\citenamefont {Ding}\ \emph {et~al.}(2020)\citenamefont {Ding}, \citenamefont {Li}, \citenamefont {Xu}, \citenamefont {Li}, \citenamefont {Hou}, \citenamefont {Liu}, \citenamefont {Xi}, \citenamefont {Xu}, \citenamefont {Yao},\ and\ \citenamefont {Wang}}]{Ding2020}%
  \BibitemOpen
  \bibfield  {author} {\bibinfo {author} {\bibfnamefont {B.}~\bibnamefont {Ding}}, \bibinfo {author} {\bibfnamefont {Z.}~\bibnamefont {Li}}, \bibinfo {author} {\bibfnamefont {G.}~\bibnamefont {Xu}}, \bibinfo {author} {\bibfnamefont {H.}~\bibnamefont {Li}}, \bibinfo {author} {\bibfnamefont {Z.}~\bibnamefont {Hou}}, \bibinfo {author} {\bibfnamefont {E.}~\bibnamefont {Liu}}, \bibinfo {author} {\bibfnamefont {X.}~\bibnamefont {Xi}}, \bibinfo {author} {\bibfnamefont {F.}~\bibnamefont {Xu}}, \bibinfo {author} {\bibfnamefont {Y.}~\bibnamefont {Yao}}, \ and\ \bibinfo {author} {\bibfnamefont {W.}~\bibnamefont {Wang}},\ }\href {\doibase 10.1021/acs.nanolett.9b03453} {\bibfield  {journal} {\bibinfo  {journal} {Nano Letters}\ }\textbf {\bibinfo {volume} {20}},\ \bibinfo {pages} {868} (\bibinfo {year} {2020})}\BibitemShut {NoStop}%
\bibitem [{\citenamefont {Meijer}\ \emph {et~al.}(2020)\citenamefont {Meijer}, \citenamefont {Lucassen}, \citenamefont {Duine}, \citenamefont {Swagten}, \citenamefont {Koopmans}, \citenamefont {Lavrijsen},\ and\ \citenamefont {Guimarães}}]{Meijer2020}%
  \BibitemOpen
  \bibfield  {author} {\bibinfo {author} {\bibfnamefont {M.~J.}\ \bibnamefont {Meijer}}, \bibinfo {author} {\bibfnamefont {J.}~\bibnamefont {Lucassen}}, \bibinfo {author} {\bibfnamefont {R.~A.}\ \bibnamefont {Duine}}, \bibinfo {author} {\bibfnamefont {H.~J.~M.}\ \bibnamefont {Swagten}}, \bibinfo {author} {\bibfnamefont {B.}~\bibnamefont {Koopmans}}, \bibinfo {author} {\bibfnamefont {R.}~\bibnamefont {Lavrijsen}}, \ and\ \bibinfo {author} {\bibfnamefont {M.~H.~D.}\ \bibnamefont {Guimarães}},\ }\href {\doibase 10.1021/acs.nanolett.0c03111} {\bibfield  {journal} {\bibinfo  {journal} {Nano Letters}\ }\textbf {\bibinfo {volume} {20}},\ \bibinfo {pages} {8563} (\bibinfo {year} {2020})}\BibitemShut {NoStop}%
\bibitem [{\citenamefont {Park}\ \emph {et~al.}(2021)\citenamefont {Park}, \citenamefont {Peng}, \citenamefont {Liang}, \citenamefont {Hallal}, \citenamefont {Yasin}, \citenamefont {Zhang}, \citenamefont {Song}, \citenamefont {Kim}, \citenamefont {Kim}, \citenamefont {Weigand}, \citenamefont {Sch\"utz}, \citenamefont {Finizio}, \citenamefont {Raabe}, \citenamefont {Garcia}, \citenamefont {Xia}, \citenamefont {Zhou}, \citenamefont {Ezawa}, \citenamefont {Liu}, \citenamefont {Chang}, \citenamefont {Koo}, \citenamefont {Kim}, \citenamefont {Chshiev}, \citenamefont {Fert}, \citenamefont {Yang}, \citenamefont {Yu},\ and\ \citenamefont {Woo}}]{Park2021}%
  \BibitemOpen
  \bibfield  {author} {\bibinfo {author} {\bibfnamefont {T.-E.}\ \bibnamefont {Park}}, \bibinfo {author} {\bibfnamefont {L.}~\bibnamefont {Peng}}, \bibinfo {author} {\bibfnamefont {J.}~\bibnamefont {Liang}}, \bibinfo {author} {\bibfnamefont {A.}~\bibnamefont {Hallal}}, \bibinfo {author} {\bibfnamefont {F.~S.}\ \bibnamefont {Yasin}}, \bibinfo {author} {\bibfnamefont {X.}~\bibnamefont {Zhang}}, \bibinfo {author} {\bibfnamefont {K.~M.}\ \bibnamefont {Song}}, \bibinfo {author} {\bibfnamefont {S.~J.}\ \bibnamefont {Kim}}, \bibinfo {author} {\bibfnamefont {K.}~\bibnamefont {Kim}}, \bibinfo {author} {\bibfnamefont {M.}~\bibnamefont {Weigand}}, \bibinfo {author} {\bibfnamefont {G.}~\bibnamefont {Sch\"utz}}, \bibinfo {author} {\bibfnamefont {S.}~\bibnamefont {Finizio}}, \bibinfo {author} {\bibfnamefont {J.}~\bibnamefont {Raabe}}, \bibinfo {author} {\bibfnamefont {K.}~\bibnamefont {Garcia}}, \bibinfo {author} {\bibfnamefont {J.}~\bibnamefont {Xia}}, \bibinfo {author} {\bibfnamefont {Y.}~\bibnamefont {Zhou}}, \bibinfo
  {author} {\bibfnamefont {M.}~\bibnamefont {Ezawa}}, \bibinfo {author} {\bibfnamefont {X.}~\bibnamefont {Liu}}, \bibinfo {author} {\bibfnamefont {J.}~\bibnamefont {Chang}}, \bibinfo {author} {\bibfnamefont {H.~C.}\ \bibnamefont {Koo}}, \bibinfo {author} {\bibfnamefont {Y.~D.}\ \bibnamefont {Kim}}, \bibinfo {author} {\bibfnamefont {M.}~\bibnamefont {Chshiev}}, \bibinfo {author} {\bibfnamefont {A.}~\bibnamefont {Fert}}, \bibinfo {author} {\bibfnamefont {H.}~\bibnamefont {Yang}}, \bibinfo {author} {\bibfnamefont {X.}~\bibnamefont {Yu}}, \ and\ \bibinfo {author} {\bibfnamefont {S.}~\bibnamefont {Woo}},\ }\href {\doibase 10.1103/PhysRevB.103.104410} {\bibfield  {journal} {\bibinfo  {journal} {Phys. Rev. B}\ }\textbf {\bibinfo {volume} {103}},\ \bibinfo {pages} {104410} (\bibinfo {year} {2021})}\BibitemShut {NoStop}%
\bibitem [{\citenamefont {Laref}\ \emph {et~al.}(2020)\citenamefont {Laref}, \citenamefont {Kim},\ and\ \citenamefont {Manchon}}]{Manchon2020}%
  \BibitemOpen
  \bibfield  {author} {\bibinfo {author} {\bibfnamefont {S.}~\bibnamefont {Laref}}, \bibinfo {author} {\bibfnamefont {K.-W.}\ \bibnamefont {Kim}}, \ and\ \bibinfo {author} {\bibfnamefont {A.}~\bibnamefont {Manchon}},\ }\href {\doibase 10.1103/PhysRevB.102.060402} {\bibfield  {journal} {\bibinfo  {journal} {Phys. Rev. B}\ }\textbf {\bibinfo {volume} {102}},\ \bibinfo {pages} {060402} (\bibinfo {year} {2020})}\BibitemShut {NoStop}%
\bibitem [{\citenamefont {Hals}\ and\ \citenamefont {Everschor-Sitte}(2019)}]{Everschor-Sitte2019}%
  \BibitemOpen
  \bibfield  {author} {\bibinfo {author} {\bibfnamefont {K.~M.~D.}\ \bibnamefont {Hals}}\ and\ \bibinfo {author} {\bibfnamefont {K.}~\bibnamefont {Everschor-Sitte}},\ }\href {\doibase 10.1103/PhysRevB.99.104422} {\bibfield  {journal} {\bibinfo  {journal} {Phys. Rev. B}\ }\textbf {\bibinfo {volume} {99}},\ \bibinfo {pages} {104422} (\bibinfo {year} {2019})}\BibitemShut {NoStop}%
\bibitem [{\citenamefont {Lifschitz}(1941)}]{Lifschitz1941}%
  \BibitemOpen
  \bibfield  {author} {\bibinfo {author} {\bibfnamefont {E.}~\bibnamefont {Lifschitz}},\ }\href@noop {} {\bibfield  {journal} {\bibinfo  {journal} {JETP}\ }\textbf {\bibinfo {volume} {11}},\ \bibinfo {pages} {253} (\bibinfo {year} {1941})}\BibitemShut {NoStop}%
\bibitem [{\citenamefont {Mitrano}\ \emph {et~al.}(2024)\citenamefont {Mitrano}, \citenamefont {Johnston}, \citenamefont {Kim},\ and\ \citenamefont {Dean}}]{Mitrano2024}%
  \BibitemOpen
  \bibfield  {author} {\bibinfo {author} {\bibfnamefont {M.}~\bibnamefont {Mitrano}}, \bibinfo {author} {\bibfnamefont {S.}~\bibnamefont {Johnston}}, \bibinfo {author} {\bibfnamefont {Y.-J.}\ \bibnamefont {Kim}}, \ and\ \bibinfo {author} {\bibfnamefont {M.~P.~M.}\ \bibnamefont {Dean}},\ }\href {\doibase 10.1103/PhysRevX.14.040501} {\bibfield  {journal} {\bibinfo  {journal} {Phys. Rev. X}\ }\textbf {\bibinfo {volume} {14}},\ \bibinfo {pages} {040501} (\bibinfo {year} {2024})}\BibitemShut {NoStop}%
\bibitem [{\citenamefont {Ament}\ \emph {et~al.}(2011)\citenamefont {Ament}, \citenamefont {van Veenendaal}, \citenamefont {Devereaux}, \citenamefont {Hill},\ and\ \citenamefont {van~den Brink}}]{RIXS}%
  \BibitemOpen
  \bibfield  {author} {\bibinfo {author} {\bibfnamefont {L.~J.~P.}\ \bibnamefont {Ament}}, \bibinfo {author} {\bibfnamefont {M.}~\bibnamefont {van Veenendaal}}, \bibinfo {author} {\bibfnamefont {T.~P.}\ \bibnamefont {Devereaux}}, \bibinfo {author} {\bibfnamefont {J.~P.}\ \bibnamefont {Hill}}, \ and\ \bibinfo {author} {\bibfnamefont {J.}~\bibnamefont {van~den Brink}},\ }\href {\doibase 10.1103/RevModPhys.83.705} {\bibfield  {journal} {\bibinfo  {journal} {Rev. Mod. Phys.}\ }\textbf {\bibinfo {volume} {83}},\ \bibinfo {pages} {705} (\bibinfo {year} {2011})}\BibitemShut {NoStop}%
\bibitem [{\citenamefont {Müller}\ \emph {et~al.}(2022)\citenamefont {Müller}, \citenamefont {Grandi},\ and\ \citenamefont {Eckstein}}]{TimeRIXS}%
  \BibitemOpen
  \bibfield  {author} {\bibinfo {author} {\bibfnamefont {A.}~\bibnamefont {Müller}}, \bibinfo {author} {\bibfnamefont {F.}~\bibnamefont {Grandi}}, \ and\ \bibinfo {author} {\bibfnamefont {M.}~\bibnamefont {Eckstein}},\ }\href {\doibase 10.1103/PhysRevB.106.L121107} {\bibfield  {journal} {\bibinfo  {journal} {Phys. Rev. B}\ }\textbf {\bibinfo {volume} {106}},\ \bibinfo {pages} {L121107} (\bibinfo {year} {2022})}\BibitemShut {NoStop}%
\bibitem [{\citenamefont {Chen}\ \emph {et~al.}(2019)\citenamefont {Chen}, \citenamefont {Wang}, \citenamefont {Jia}, \citenamefont {Moritz}, \citenamefont {Shvaika}, \citenamefont {Freericks},\ and\ \citenamefont {Devereaux}}]{TimeRIXS2}%
  \BibitemOpen
  \bibfield  {author} {\bibinfo {author} {\bibfnamefont {Y.}~\bibnamefont {Chen}}, \bibinfo {author} {\bibfnamefont {Y.}~\bibnamefont {Wang}}, \bibinfo {author} {\bibfnamefont {C.}~\bibnamefont {Jia}}, \bibinfo {author} {\bibfnamefont {B.}~\bibnamefont {Moritz}}, \bibinfo {author} {\bibfnamefont {A.~M.}\ \bibnamefont {Shvaika}}, \bibinfo {author} {\bibfnamefont {J.~K.}\ \bibnamefont {Freericks}}, \ and\ \bibinfo {author} {\bibfnamefont {T.~P.}\ \bibnamefont {Devereaux}},\ }\href {\doibase 10.1103/PhysRevB.99.104306} {\bibfield  {journal} {\bibinfo  {journal} {Phys. Rev. B}\ }\textbf {\bibinfo {volume} {99}},\ \bibinfo {pages} {104306} (\bibinfo {year} {2019})}\BibitemShut {NoStop}%
\bibitem [{\citenamefont {Dzyaloshinsky}(1958)}]{Dzyaloshinsky1958}%
  \BibitemOpen
  \bibfield  {author} {\bibinfo {author} {\bibfnamefont {I.}~\bibnamefont {Dzyaloshinsky}},\ }\href@noop {} {\bibfield  {journal} {\bibinfo  {journal} {Journal of Physics and Chemistry of Solids}\ }\textbf {\bibinfo {volume} {4}},\ \bibinfo {pages} {241} (\bibinfo {year} {1958})}\BibitemShut {NoStop}%
\bibitem [{\citenamefont {Moriya}(1960)}]{Moriya1960}%
  \BibitemOpen
  \bibfield  {author} {\bibinfo {author} {\bibfnamefont {T.}~\bibnamefont {Moriya}},\ }\href {\doibase 10.1103/PhysRev.120.91} {\bibfield  {journal} {\bibinfo  {journal} {Phys. Rev.}\ }\textbf {\bibinfo {volume} {120}},\ \bibinfo {pages} {91} (\bibinfo {year} {1960})}\BibitemShut {NoStop}%
\bibitem [{\citenamefont {Kuepferling}\ \emph {et~al.}(2023)\citenamefont {Kuepferling}, \citenamefont {Casiraghi}, \citenamefont {Soares}, \citenamefont {Durin}, \citenamefont {Garcia-Sanchez}, \citenamefont {Chen}, \citenamefont {Back}, \citenamefont {Marrows}, \citenamefont {Tacchi},\ and\ \citenamefont {Carlotti}}]{Kuepferling2023}%
  \BibitemOpen
  \bibfield  {author} {\bibinfo {author} {\bibfnamefont {M.}~\bibnamefont {Kuepferling}}, \bibinfo {author} {\bibfnamefont {A.}~\bibnamefont {Casiraghi}}, \bibinfo {author} {\bibfnamefont {G.}~\bibnamefont {Soares}}, \bibinfo {author} {\bibfnamefont {G.}~\bibnamefont {Durin}}, \bibinfo {author} {\bibfnamefont {F.}~\bibnamefont {Garcia-Sanchez}}, \bibinfo {author} {\bibfnamefont {L.}~\bibnamefont {Chen}}, \bibinfo {author} {\bibfnamefont {C.~H.}\ \bibnamefont {Back}}, \bibinfo {author} {\bibfnamefont {C.~H.}\ \bibnamefont {Marrows}}, \bibinfo {author} {\bibfnamefont {S.}~\bibnamefont {Tacchi}}, \ and\ \bibinfo {author} {\bibfnamefont {G.}~\bibnamefont {Carlotti}},\ }\href {\doibase 10.1103/RevModPhys.95.015003} {\bibfield  {journal} {\bibinfo  {journal} {Rev. Mod. Phys.}\ }\textbf {\bibinfo {volume} {95}},\ \bibinfo {pages} {015003} (\bibinfo {year} {2023})}\BibitemShut {NoStop}%
\bibitem [{\citenamefont {Liang}\ \emph {et~al.}(2020)\citenamefont {Liang}, \citenamefont {Wang}, \citenamefont {Du}, \citenamefont {Hallal}, \citenamefont {Garcia}, \citenamefont {Chshiev}, \citenamefont {Fert},\ and\ \citenamefont {Yang}}]{Liang2020}%
  \BibitemOpen
  \bibfield  {author} {\bibinfo {author} {\bibfnamefont {J.}~\bibnamefont {Liang}}, \bibinfo {author} {\bibfnamefont {W.}~\bibnamefont {Wang}}, \bibinfo {author} {\bibfnamefont {H.}~\bibnamefont {Du}}, \bibinfo {author} {\bibfnamefont {A.}~\bibnamefont {Hallal}}, \bibinfo {author} {\bibfnamefont {K.}~\bibnamefont {Garcia}}, \bibinfo {author} {\bibfnamefont {M.}~\bibnamefont {Chshiev}}, \bibinfo {author} {\bibfnamefont {A.}~\bibnamefont {Fert}}, \ and\ \bibinfo {author} {\bibfnamefont {H.}~\bibnamefont {Yang}},\ }\href {\doibase 10.1103/PhysRevB.101.184401} {\bibfield  {journal} {\bibinfo  {journal} {Phys. Rev. B}\ }\textbf {\bibinfo {volume} {101}},\ \bibinfo {pages} {184401} (\bibinfo {year} {2020})}\BibitemShut {NoStop}%
\bibitem [{\citenamefont {Mazurenko}\ \emph {et~al.}(2021)\citenamefont {Mazurenko}, \citenamefont {Kvashnin}, \citenamefont {Lichtenstein},\ and\ \citenamefont {Katsnelson}}]{Mazurenko2021}%
  \BibitemOpen
  \bibfield  {author} {\bibinfo {author} {\bibfnamefont {V.~V.}\ \bibnamefont {Mazurenko}}, \bibinfo {author} {\bibfnamefont {Y.~O.}\ \bibnamefont {Kvashnin}}, \bibinfo {author} {\bibfnamefont {A.~I.}\ \bibnamefont {Lichtenstein}}, \ and\ \bibinfo {author} {\bibfnamefont {M.~I.}\ \bibnamefont {Katsnelson}},\ }\href@noop {} {\bibfield  {journal} {\bibinfo  {journal} {Journal of Experimental and Theoretical Physics}\ }\textbf {\bibinfo {volume} {132}},\ \bibinfo {pages} {506} (\bibinfo {year} {2021})}\BibitemShut {NoStop}%
\bibitem [{\citenamefont {Yang}\ \emph {et~al.}(2020)\citenamefont {Yang}, \citenamefont {Pellatz}, \citenamefont {Wolf}, \citenamefont {Nandkishore},\ and\ \citenamefont {Reznik}}]{Reznik2020}%
  \BibitemOpen
  \bibfield  {author} {\bibinfo {author} {\bibfnamefont {J.-A.}\ \bibnamefont {Yang}}, \bibinfo {author} {\bibfnamefont {N.}~\bibnamefont {Pellatz}}, \bibinfo {author} {\bibfnamefont {T.}~\bibnamefont {Wolf}}, \bibinfo {author} {\bibfnamefont {R.}~\bibnamefont {Nandkishore}}, \ and\ \bibinfo {author} {\bibfnamefont {D.}~\bibnamefont {Reznik}},\ }\href {\doibase 10.1038/s41467-020-16275-9} {\bibfield  {journal} {\bibinfo  {journal} {Nature Communications}\ }\textbf {\bibinfo {volume} {11}},\ \bibinfo {pages} {2548} (\bibinfo {year} {2020})}\BibitemShut {NoStop}%
\bibitem [{\citenamefont {Da{\.a}browski}\ \emph {et~al.}(2022)\citenamefont {Da{\.a}browski}, \citenamefont {Guo}, \citenamefont {Strungaru}, \citenamefont {Keatley}, \citenamefont {Withers}, \citenamefont {Santos},\ and\ \citenamefont {Hicken}}]{Dabrowski2022}%
  \BibitemOpen
  \bibfield  {author} {\bibinfo {author} {\bibfnamefont {M.}~\bibnamefont {Da{\.a}browski}}, \bibinfo {author} {\bibfnamefont {S.}~\bibnamefont {Guo}}, \bibinfo {author} {\bibfnamefont {M.}~\bibnamefont {Strungaru}}, \bibinfo {author} {\bibfnamefont {P.~S.}\ \bibnamefont {Keatley}}, \bibinfo {author} {\bibfnamefont {F.}~\bibnamefont {Withers}}, \bibinfo {author} {\bibfnamefont {E.~J.~G.}\ \bibnamefont {Santos}}, \ and\ \bibinfo {author} {\bibfnamefont {R.~J.}\ \bibnamefont {Hicken}},\ }\href {\doibase 10.1038/s41467-022-33343-4} {\bibfield  {journal} {\bibinfo  {journal} {Nature Communications}\ }\textbf {\bibinfo {volume} {13}},\ \bibinfo {pages} {5976} (\bibinfo {year} {2022})}\BibitemShut {NoStop}%
\bibitem [{\citenamefont {Vahaplar}\ \emph {et~al.}(2009)\citenamefont {Vahaplar}, \citenamefont {Kalashnikova}, \citenamefont {Kimel}, \citenamefont {Hinzke}, \citenamefont {Nowak}, \citenamefont {Chantrell}, \citenamefont {Tsukamoto}, \citenamefont {Itoh}, \citenamefont {Kirilyuk},\ and\ \citenamefont {Rasing}}]{Vahaplar2009}%
  \BibitemOpen
  \bibfield  {author} {\bibinfo {author} {\bibfnamefont {K.}~\bibnamefont {Vahaplar}}, \bibinfo {author} {\bibfnamefont {A.~M.}\ \bibnamefont {Kalashnikova}}, \bibinfo {author} {\bibfnamefont {A.~V.}\ \bibnamefont {Kimel}}, \bibinfo {author} {\bibfnamefont {D.}~\bibnamefont {Hinzke}}, \bibinfo {author} {\bibfnamefont {U.}~\bibnamefont {Nowak}}, \bibinfo {author} {\bibfnamefont {R.}~\bibnamefont {Chantrell}}, \bibinfo {author} {\bibfnamefont {A.}~\bibnamefont {Tsukamoto}}, \bibinfo {author} {\bibfnamefont {A.}~\bibnamefont {Itoh}}, \bibinfo {author} {\bibfnamefont {A.}~\bibnamefont {Kirilyuk}}, \ and\ \bibinfo {author} {\bibfnamefont {T.}~\bibnamefont {Rasing}},\ }\href {\doibase 10.1103/PhysRevLett.103.117201} {\bibfield  {journal} {\bibinfo  {journal} {Phys. Rev. Lett.}\ }\textbf {\bibinfo {volume} {103}},\ \bibinfo {pages} {117201} (\bibinfo {year} {2009})}\BibitemShut {NoStop}%
\bibitem [{\citenamefont {Radu}\ \emph {et~al.}(2011)\citenamefont {Radu}, \citenamefont {Vahaplar}, \citenamefont {Stamm}, \citenamefont {Kachel}, \citenamefont {Pontius}, \citenamefont {Dürr}, \citenamefont {Ostler}, \citenamefont {Barker}, \citenamefont {Evans}, \citenamefont {Chantrell}, \citenamefont {Tsukamoto}, \citenamefont {Itoh}, \citenamefont {Kirilyuk}, \citenamefont {Rasing},\ and\ \citenamefont {Kimel}}]{Radu2011}%
  \BibitemOpen
  \bibfield  {author} {\bibinfo {author} {\bibfnamefont {I.}~\bibnamefont {Radu}}, \bibinfo {author} {\bibfnamefont {K.}~\bibnamefont {Vahaplar}}, \bibinfo {author} {\bibfnamefont {C.}~\bibnamefont {Stamm}}, \bibinfo {author} {\bibfnamefont {T.}~\bibnamefont {Kachel}}, \bibinfo {author} {\bibfnamefont {N.}~\bibnamefont {Pontius}}, \bibinfo {author} {\bibfnamefont {H.~A.}\ \bibnamefont {Dürr}}, \bibinfo {author} {\bibfnamefont {T.~A.}\ \bibnamefont {Ostler}}, \bibinfo {author} {\bibfnamefont {J.}~\bibnamefont {Barker}}, \bibinfo {author} {\bibfnamefont {R.~F.~L.}\ \bibnamefont {Evans}}, \bibinfo {author} {\bibfnamefont {R.~W.}\ \bibnamefont {Chantrell}}, \bibinfo {author} {\bibfnamefont {A.}~\bibnamefont {Tsukamoto}}, \bibinfo {author} {\bibfnamefont {A.}~\bibnamefont {Itoh}}, \bibinfo {author} {\bibfnamefont {A.}~\bibnamefont {Kirilyuk}}, \bibinfo {author} {\bibfnamefont {T.}~\bibnamefont {Rasing}}, \ and\ \bibinfo {author} {\bibfnamefont {A.~V.}\ \bibnamefont {Kimel}},\ }\href {\doibase 10.1038/nature09901}
  {\bibfield  {journal} {\bibinfo  {journal} {Nature}\ }\textbf {\bibinfo {volume} {472}},\ \bibinfo {pages} {205} (\bibinfo {year} {2011})}\BibitemShut {NoStop}%
\bibitem [{\citenamefont {Keldysh}(1965)}]{Keldysh}%
  \BibitemOpen
  \bibfield  {author} {\bibinfo {author} {\bibfnamefont {L.~V.}\ \bibnamefont {Keldysh}},\ }\href@noop {} {\bibfield  {journal} {\bibinfo  {journal} {Sov. Phys. JETP}\ }\textbf {\bibinfo {volume} {20}},\ \bibinfo {pages} {1018} (\bibinfo {year} {1965})}\BibitemShut {NoStop}%
\bibitem [{\citenamefont {Rammer}\ and\ \citenamefont {Smith}(1986)}]{Rammer}%
  \BibitemOpen
  \bibfield  {author} {\bibinfo {author} {\bibfnamefont {J.}~\bibnamefont {Rammer}}\ and\ \bibinfo {author} {\bibfnamefont {H.}~\bibnamefont {Smith}},\ }\href@noop {} {\bibfield  {journal} {\bibinfo  {journal} {Rev. Mod. Phys.}\ }\textbf {\bibinfo {volume} {58}},\ \bibinfo {pages} {323} (\bibinfo {year} {1986})}\BibitemShut {NoStop}%
\bibitem [{\citenamefont {Kamenev}(2011)}]{Kamenev}%
  \BibitemOpen
  \bibfield  {author} {\bibinfo {author} {\bibfnamefont {A.}~\bibnamefont {Kamenev}},\ }\href@noop {} {\emph {\bibinfo {title} {Field Theory of Non-Equilibrium Systems}}}\ (\bibinfo  {publisher} {Cambridge University Press},\ \bibinfo {year} {2011})\BibitemShut {NoStop}%
\bibitem [{\citenamefont {Lifshitz}(1995)}]{Lifshitz}%
  \BibitemOpen
  \bibfield  {author} {\bibinfo {author} {\bibfnamefont {L.~P. P. E.~M.}\ \bibnamefont {Lifshitz}},\ }\href@noop {} {\emph {\bibinfo {title} {Physical kinetics: Volume 10}}}\ (\bibinfo  {publisher} {Elsevier},\ \bibinfo {address} {Science},\ \bibinfo {year} {1995})\BibitemShut {NoStop}%
\bibitem [{\citenamefont {Calzetta}\ and\ \citenamefont {Hu.}(2008)}]{Calzetta}%
  \BibitemOpen
  \bibfield  {author} {\bibinfo {author} {\bibfnamefont {E.~A.}\ \bibnamefont {Calzetta}}\ and\ \bibinfo {author} {\bibfnamefont {B.~B.}\ \bibnamefont {Hu.}},\ }\href@noop {} {\emph {\bibinfo {title} {Nonequilibrium Quantum Field Theory}}}\ (\bibinfo  {publisher} {Cambridge Monographs on Mathematical Physics. Cambridge University Press},\ \bibinfo {address} {9},\ \bibinfo {year} {2008})\BibitemShut {NoStop}%
\bibitem [{\citenamefont {Beaurepaire}\ \emph {et~al.}(1996{\natexlab{a}})\citenamefont {Beaurepaire}, \citenamefont {Merle}, \citenamefont {Daunois},\ and\ \citenamefont {Bigot}}]{Beaurepaire1996}%
  \BibitemOpen
  \bibfield  {author} {\bibinfo {author} {\bibfnamefont {E.}~\bibnamefont {Beaurepaire}}, \bibinfo {author} {\bibfnamefont {J.-C.}\ \bibnamefont {Merle}}, \bibinfo {author} {\bibfnamefont {A.}~\bibnamefont {Daunois}}, \ and\ \bibinfo {author} {\bibfnamefont {J.-Y.}\ \bibnamefont {Bigot}},\ }\href {\doibase 10.1103/PhysRevLett.76.4250} {\bibfield  {journal} {\bibinfo  {journal} {Phys. Rev. Lett.}\ }\textbf {\bibinfo {volume} {76}},\ \bibinfo {pages} {4250} (\bibinfo {year} {1996}{\natexlab{a}})}\BibitemShut {NoStop}%
\bibitem [{\citenamefont {Shah}(1999)}]{Shah1999}%
  \BibitemOpen
  \bibfield  {author} {\bibinfo {author} {\bibfnamefont {J.~J.}\ \bibnamefont {Shah}},\ }\href@noop {} {\emph {\bibinfo {title} {Ultrafast spectroscopy of semiconductors and semiconductor nanostructures}}},\ \bibinfo {edition} {2nd}\ ed.,\ edited by\ \bibinfo {editor} {\bibfnamefont {B.~L.}\ \bibnamefont {Innovations}},\ Springer series in solid-state sciences\ (\bibinfo  {publisher} {Springer},\ \bibinfo {address} {Berlin ; New York},\ \bibinfo {year} {1999})\ p.\ \bibinfo {pages} {518},\ \bibinfo {note} {"Lucent Technologies, Bell Labs Innovations."}\BibitemShut {NoStop}%
\bibitem [{\citenamefont {Qiu}\ \emph {et~al.}(2021)\citenamefont {Qiu}, \citenamefont {Zhou}, \citenamefont {Zhang}, \citenamefont {Wu}, \citenamefont {Tian}, \citenamefont {Cheng}, \citenamefont {Mi}, \citenamefont {Zhao}, \citenamefont {Zhang}, \citenamefont {Wu}, \citenamefont {Jin}, \citenamefont {Chen},\ and\ \citenamefont {Wu}}]{Qiu2021}%
  \BibitemOpen
  \bibfield  {author} {\bibinfo {author} {\bibfnamefont {H.}~\bibnamefont {Qiu}}, \bibinfo {author} {\bibfnamefont {L.}~\bibnamefont {Zhou}}, \bibinfo {author} {\bibfnamefont {C.}~\bibnamefont {Zhang}}, \bibinfo {author} {\bibfnamefont {J.}~\bibnamefont {Wu}}, \bibinfo {author} {\bibfnamefont {Y.}~\bibnamefont {Tian}}, \bibinfo {author} {\bibfnamefont {S.}~\bibnamefont {Cheng}}, \bibinfo {author} {\bibfnamefont {S.}~\bibnamefont {Mi}}, \bibinfo {author} {\bibfnamefont {H.}~\bibnamefont {Zhao}}, \bibinfo {author} {\bibfnamefont {Q.}~\bibnamefont {Zhang}}, \bibinfo {author} {\bibfnamefont {D.}~\bibnamefont {Wu}}, \bibinfo {author} {\bibfnamefont {B.}~\bibnamefont {Jin}}, \bibinfo {author} {\bibfnamefont {J.}~\bibnamefont {Chen}}, \ and\ \bibinfo {author} {\bibfnamefont {P.}~\bibnamefont {Wu}},\ }\href {\doibase 10.1038/s41567-020-01061-7} {\bibfield  {journal} {\bibinfo  {journal} {Nature Physics}\ }\textbf {\bibinfo {volume} {17}},\ \bibinfo {pages} {388} (\bibinfo {year} {2021})}\BibitemShut {NoStop}%
\bibitem [{\citenamefont {Hennecke}\ \emph {et~al.}(2019)\citenamefont {Hennecke}, \citenamefont {Radu}, \citenamefont {Abrudan}, \citenamefont {Kachel}, \citenamefont {Holldack}, \citenamefont {Mitzner}, \citenamefont {Tsukamoto},\ and\ \citenamefont {Eisebitt}}]{Hennecke}%
  \BibitemOpen
  \bibfield  {author} {\bibinfo {author} {\bibfnamefont {M.}~\bibnamefont {Hennecke}}, \bibinfo {author} {\bibfnamefont {I.}~\bibnamefont {Radu}}, \bibinfo {author} {\bibfnamefont {R.}~\bibnamefont {Abrudan}}, \bibinfo {author} {\bibfnamefont {T.}~\bibnamefont {Kachel}}, \bibinfo {author} {\bibfnamefont {K.}~\bibnamefont {Holldack}}, \bibinfo {author} {\bibfnamefont {R.}~\bibnamefont {Mitzner}}, \bibinfo {author} {\bibfnamefont {A.}~\bibnamefont {Tsukamoto}}, \ and\ \bibinfo {author} {\bibfnamefont {S.}~\bibnamefont {Eisebitt}},\ }\href {\doibase 10.1103/PhysRevLett.122.157202} {\bibfield  {journal} {\bibinfo  {journal} {Phys. Rev. Lett.}\ }\textbf {\bibinfo {volume} {122}},\ \bibinfo {pages} {157202} (\bibinfo {year} {2019})}\BibitemShut {NoStop}%
\bibitem [{\citenamefont {Pogrebna}\ \emph {et~al.}(2019)\citenamefont {Pogrebna}, \citenamefont {Prabhakara}, \citenamefont {Davydova}, \citenamefont {Becker}, \citenamefont {Tsukamoto}, \citenamefont {Rasing}, \citenamefont {Kirilyuk}, \citenamefont {Zvezdin}, \citenamefont {Christianen},\ and\ \citenamefont {Kimel}}]{Pogrebna}%
  \BibitemOpen
  \bibfield  {author} {\bibinfo {author} {\bibfnamefont {A.}~\bibnamefont {Pogrebna}}, \bibinfo {author} {\bibfnamefont {K.}~\bibnamefont {Prabhakara}}, \bibinfo {author} {\bibfnamefont {M.}~\bibnamefont {Davydova}}, \bibinfo {author} {\bibfnamefont {J.}~\bibnamefont {Becker}}, \bibinfo {author} {\bibfnamefont {A.}~\bibnamefont {Tsukamoto}}, \bibinfo {author} {\bibfnamefont {T.}~\bibnamefont {Rasing}}, \bibinfo {author} {\bibfnamefont {A.}~\bibnamefont {Kirilyuk}}, \bibinfo {author} {\bibfnamefont {A.~K.}\ \bibnamefont {Zvezdin}}, \bibinfo {author} {\bibfnamefont {P.~C.~M.}\ \bibnamefont {Christianen}}, \ and\ \bibinfo {author} {\bibfnamefont {A.}~\bibnamefont {Kimel}},\ }\href {\doibase 10.1103/PhysRevB.100.174427} {\bibfield  {journal} {\bibinfo  {journal} {Phys. Rev. B}\ }\textbf {\bibinfo {volume} {100}},\ \bibinfo {pages} {174427} (\bibinfo {year} {2019})}\BibitemShut {NoStop}%
\bibitem [{\citenamefont {Mikhaylovskiy}\ \emph {et~al.}(2015)\citenamefont {Mikhaylovskiy}, \citenamefont {Hendry}, \citenamefont {Secchi}, \citenamefont {Mentink}, \citenamefont {Eckstein}, \citenamefont {Wu}, \citenamefont {Pisarev}, \citenamefont {Kruglyak}, \citenamefont {Katsnelson}, \citenamefont {Rasing},\ and\ \citenamefont {Kimel}}]{FeO}%
  \BibitemOpen
  \bibfield  {author} {\bibinfo {author} {\bibfnamefont {R.~V.}\ \bibnamefont {Mikhaylovskiy}}, \bibinfo {author} {\bibfnamefont {E.}~\bibnamefont {Hendry}}, \bibinfo {author} {\bibfnamefont {A.}~\bibnamefont {Secchi}}, \bibinfo {author} {\bibfnamefont {J.~H.}\ \bibnamefont {Mentink}}, \bibinfo {author} {\bibfnamefont {M.}~\bibnamefont {Eckstein}}, \bibinfo {author} {\bibfnamefont {A.}~\bibnamefont {Wu}}, \bibinfo {author} {\bibfnamefont {R.~V.}\ \bibnamefont {Pisarev}}, \bibinfo {author} {\bibfnamefont {V.~V.}\ \bibnamefont {Kruglyak}}, \bibinfo {author} {\bibfnamefont {M.~I.}\ \bibnamefont {Katsnelson}}, \bibinfo {author} {\bibfnamefont {T.}~\bibnamefont {Rasing}}, \ and\ \bibinfo {author} {\bibfnamefont {A.~V.}\ \bibnamefont {Kimel}},\ }\href {\doibase 10.1038/ncomms9190} {\bibfield  {journal} {\bibinfo  {journal} {Nature Communications}\ }\textbf {\bibinfo {volume} {6}},\ \bibinfo {pages} {8190} (\bibinfo {year} {2015})}\BibitemShut {NoStop}%
\bibitem [{\citenamefont {Beaurepaire}\ \emph {et~al.}(1996{\natexlab{b}})\citenamefont {Beaurepaire}, \citenamefont {Merle}, \citenamefont {Daunois},\ and\ \citenamefont {Bigot}}]{Bigot2005}%
  \BibitemOpen
  \bibfield  {author} {\bibinfo {author} {\bibfnamefont {E.}~\bibnamefont {Beaurepaire}}, \bibinfo {author} {\bibfnamefont {J.-C.}\ \bibnamefont {Merle}}, \bibinfo {author} {\bibfnamefont {A.}~\bibnamefont {Daunois}}, \ and\ \bibinfo {author} {\bibfnamefont {J.-Y.}\ \bibnamefont {Bigot}},\ }\href {\doibase 10.1103/PhysRevLett.76.4250} {\bibfield  {journal} {\bibinfo  {journal} {Phys. Rev. Lett.}\ }\textbf {\bibinfo {volume} {76}},\ \bibinfo {pages} {4250} (\bibinfo {year} {1996}{\natexlab{b}})}\BibitemShut {NoStop}%
\bibitem [{\citenamefont {Kirilyuk}\ \emph {et~al.}(2010)\citenamefont {Kirilyuk}, \citenamefont {Kimel},\ and\ \citenamefont {Rasing}}]{RasingRMP}%
  \BibitemOpen
  \bibfield  {author} {\bibinfo {author} {\bibfnamefont {A.}~\bibnamefont {Kirilyuk}}, \bibinfo {author} {\bibfnamefont {A.~V.}\ \bibnamefont {Kimel}}, \ and\ \bibinfo {author} {\bibfnamefont {T.}~\bibnamefont {Rasing}},\ }\href {\doibase 10.1103/RevModPhys.82.2731} {\bibfield  {journal} {\bibinfo  {journal} {Rev. Mod. Phys.}\ }\textbf {\bibinfo {volume} {82}},\ \bibinfo {pages} {2731} (\bibinfo {year} {2010})}\BibitemShut {NoStop}%
\bibitem [{\citenamefont {Subkhangulov}\ \emph {et~al.}(2014)\citenamefont {Subkhangulov}, \citenamefont {Henriques}, \citenamefont {Rappl}, \citenamefont {Abramof}, \citenamefont {Rasing},\ and\ \citenamefont {Kimel}}]{Rasing2014}%
  \BibitemOpen
  \bibfield  {author} {\bibinfo {author} {\bibfnamefont {R.~R.}\ \bibnamefont {Subkhangulov}}, \bibinfo {author} {\bibfnamefont {A.~B.}\ \bibnamefont {Henriques}}, \bibinfo {author} {\bibfnamefont {P.~H.~O.}\ \bibnamefont {Rappl}}, \bibinfo {author} {\bibfnamefont {E.}~\bibnamefont {Abramof}}, \bibinfo {author} {\bibfnamefont {T.}~\bibnamefont {Rasing}}, \ and\ \bibinfo {author} {\bibfnamefont {A.~V.}\ \bibnamefont {Kimel}},\ }\href {\doibase 10.1038/srep04368} {\bibfield  {journal} {\bibinfo  {journal} {Scientific Reports}\ }\textbf {\bibinfo {volume} {4}},\ \bibinfo {pages} {4368} (\bibinfo {year} {2014})}\BibitemShut {NoStop}%
\bibitem [{\citenamefont {Kimel}\ \emph {et~al.}(2005)\citenamefont {Kimel}, \citenamefont {Kirilyuk}, \citenamefont {Usachev}, \citenamefont {Pisarev}, \citenamefont {Balbashov},\ and\ \citenamefont {Rasing}}]{Rasing2005}%
  \BibitemOpen
  \bibfield  {author} {\bibinfo {author} {\bibfnamefont {A.~V.}\ \bibnamefont {Kimel}}, \bibinfo {author} {\bibfnamefont {A.}~\bibnamefont {Kirilyuk}}, \bibinfo {author} {\bibfnamefont {P.~A.}\ \bibnamefont {Usachev}}, \bibinfo {author} {\bibfnamefont {R.~V.}\ \bibnamefont {Pisarev}}, \bibinfo {author} {\bibfnamefont {A.~M.}\ \bibnamefont {Balbashov}}, \ and\ \bibinfo {author} {\bibfnamefont {T.}~\bibnamefont {Rasing}},\ }\href {\doibase 10.1038/nature03564} {\bibfield  {journal} {\bibinfo  {journal} {Nature}\ }\textbf {\bibinfo {volume} {435}},\ \bibinfo {pages} {655} (\bibinfo {year} {2005})}\BibitemShut {NoStop}%
\bibitem [{\citenamefont {Atxitia}\ and\ \citenamefont {Chubykalo-Fesenko}(2011)}]{Atxitia}%
  \BibitemOpen
  \bibfield  {author} {\bibinfo {author} {\bibfnamefont {U.}~\bibnamefont {Atxitia}}\ and\ \bibinfo {author} {\bibfnamefont {O.}~\bibnamefont {Chubykalo-Fesenko}},\ }\href {\doibase 10.1103/PhysRevB.84.144414} {\bibfield  {journal} {\bibinfo  {journal} {Phys. Rev. B}\ }\textbf {\bibinfo {volume} {84}},\ \bibinfo {pages} {144414} (\bibinfo {year} {2011})}\BibitemShut {NoStop}%
\bibitem [{\citenamefont {Mendil}\ \emph {et~al.}(2014)\citenamefont {Mendil}, \citenamefont {Nieves}, \citenamefont {Chubykalo-Fesenko}, \citenamefont {Walowski}, \citenamefont {Santos}, \citenamefont {Pisana},\ and\ \citenamefont {M{\"u}nzenberg}}]{Mendil}%
  \BibitemOpen
  \bibfield  {author} {\bibinfo {author} {\bibfnamefont {J.}~\bibnamefont {Mendil}}, \bibinfo {author} {\bibfnamefont {P.}~\bibnamefont {Nieves}}, \bibinfo {author} {\bibfnamefont {O.}~\bibnamefont {Chubykalo-Fesenko}}, \bibinfo {author} {\bibfnamefont {J.}~\bibnamefont {Walowski}}, \bibinfo {author} {\bibfnamefont {T.}~\bibnamefont {Santos}}, \bibinfo {author} {\bibfnamefont {S.}~\bibnamefont {Pisana}}, \ and\ \bibinfo {author} {\bibfnamefont {M.}~\bibnamefont {M{\"u}nzenberg}},\ }\href@noop {} {\bibfield  {journal} {\bibinfo  {journal} {Scientific Reports}\ }\textbf {\bibinfo {volume} {4(1)}},\ \bibinfo {pages} {3980} (\bibinfo {year} {2014})}\BibitemShut {NoStop}%
\bibitem [{\citenamefont {Mentink}\ \emph {et~al.}(2012)\citenamefont {Mentink}, \citenamefont {Hellsvik}, \citenamefont {Afanasiev}, \citenamefont {Ivanov}, \citenamefont {Kirilyuk}, \citenamefont {Kimel}, \citenamefont {Eriksson}, \citenamefont {Katsnelson},\ and\ \citenamefont {Rasing}}]{Mentink2012}%
  \BibitemOpen
  \bibfield  {author} {\bibinfo {author} {\bibfnamefont {J.~H.}\ \bibnamefont {Mentink}}, \bibinfo {author} {\bibfnamefont {J.}~\bibnamefont {Hellsvik}}, \bibinfo {author} {\bibfnamefont {D.~V.}\ \bibnamefont {Afanasiev}}, \bibinfo {author} {\bibfnamefont {B.~A.}\ \bibnamefont {Ivanov}}, \bibinfo {author} {\bibfnamefont {A.}~\bibnamefont {Kirilyuk}}, \bibinfo {author} {\bibfnamefont {A.~V.}\ \bibnamefont {Kimel}}, \bibinfo {author} {\bibfnamefont {O.}~\bibnamefont {Eriksson}}, \bibinfo {author} {\bibfnamefont {M.~I.}\ \bibnamefont {Katsnelson}}, \ and\ \bibinfo {author} {\bibfnamefont {T.}~\bibnamefont {Rasing}},\ }\href@noop {} {\bibfield  {journal} {\bibinfo  {journal} {Phys. Rev. Lett.}\ }\textbf {\bibinfo {volume} {108}},\ \bibinfo {pages} {057202} (\bibinfo {year} {2012})}\BibitemShut {NoStop}%
\bibitem [{\citenamefont {Mentink}\ \emph {et~al.}(2015)\citenamefont {Mentink}, \citenamefont {Balzer},\ and\ \citenamefont {Eckstein}}]{Mentink2015}%
  \BibitemOpen
  \bibfield  {author} {\bibinfo {author} {\bibfnamefont {J.~H.}\ \bibnamefont {Mentink}}, \bibinfo {author} {\bibfnamefont {K.}~\bibnamefont {Balzer}}, \ and\ \bibinfo {author} {\bibfnamefont {M.}~\bibnamefont {Eckstein}},\ }\href@noop {} {\bibfield  {journal} {\bibinfo  {journal} {Nat. Commun.}\ }\textbf {\bibinfo {volume} {6}},\ \bibinfo {pages} {6708} (\bibinfo {year} {2015})}\BibitemShut {NoStop}%
\bibitem [{\citenamefont {Itin}\ and\ \citenamefont {Katsnelson}(2015)}]{Itin2015}%
  \BibitemOpen
  \bibfield  {author} {\bibinfo {author} {\bibfnamefont {A.~P.}\ \bibnamefont {Itin}}\ and\ \bibinfo {author} {\bibfnamefont {M.~I.}\ \bibnamefont {Katsnelson}},\ }\href@noop {} {\bibfield  {journal} {\bibinfo  {journal} {Phys. Rev. Lett.}\ }\textbf {\bibinfo {volume} {115}},\ \bibinfo {pages} {075301} (\bibinfo {year} {2015})}\BibitemShut {NoStop}%
\bibitem [{\citenamefont {Dolgikh}\ \emph {et~al.}(2023)\citenamefont {Dolgikh}, \citenamefont {Afanasiev}, \citenamefont {Yurlov}, \citenamefont {Logunov}, \citenamefont {Zvezdin},\ and\ \citenamefont {Kimel}}]{Dolgikh23}%
  \BibitemOpen
  \bibfield  {author} {\bibinfo {author} {\bibfnamefont {A.}~\bibnamefont {Dolgikh}}, \bibinfo {author} {\bibfnamefont {D.}~\bibnamefont {Afanasiev}}, \bibinfo {author} {\bibfnamefont {V.~V.}\ \bibnamefont {Yurlov}}, \bibinfo {author} {\bibfnamefont {M.~V.}\ \bibnamefont {Logunov}}, \bibinfo {author} {\bibfnamefont {A.~K.}\ \bibnamefont {Zvezdin}}, \ and\ \bibinfo {author} {\bibfnamefont {A.~V.}\ \bibnamefont {Kimel}},\ }\href {\doibase 10.1103/PhysRevB.107.094424} {\bibfield  {journal} {\bibinfo  {journal} {Phys. Rev. B}\ }\textbf {\bibinfo {volume} {107}},\ \bibinfo {pages} {094424} (\bibinfo {year} {2023})}\BibitemShut {NoStop}%
\bibitem [{\citenamefont {Davies}\ \emph {et~al.}(2020)\citenamefont {Davies}, \citenamefont {Bonfiglio}, \citenamefont {Rode}, \citenamefont {Besbas}, \citenamefont {Banerjee}, \citenamefont {Stamenov}, \citenamefont {Coey}, \citenamefont {Kimel},\ and\ \citenamefont {Kirilyuk}}]{Davies20}%
  \BibitemOpen
  \bibfield  {author} {\bibinfo {author} {\bibfnamefont {C.~S.}\ \bibnamefont {Davies}}, \bibinfo {author} {\bibfnamefont {G.}~\bibnamefont {Bonfiglio}}, \bibinfo {author} {\bibfnamefont {K.}~\bibnamefont {Rode}}, \bibinfo {author} {\bibfnamefont {J.}~\bibnamefont {Besbas}}, \bibinfo {author} {\bibfnamefont {C.}~\bibnamefont {Banerjee}}, \bibinfo {author} {\bibfnamefont {P.}~\bibnamefont {Stamenov}}, \bibinfo {author} {\bibfnamefont {J.~M.~D.}\ \bibnamefont {Coey}}, \bibinfo {author} {\bibfnamefont {A.~V.}\ \bibnamefont {Kimel}}, \ and\ \bibinfo {author} {\bibfnamefont {A.}~\bibnamefont {Kirilyuk}},\ }\href {\doibase 10.1103/PhysRevResearch.2.032044} {\bibfield  {journal} {\bibinfo  {journal} {Phys. Rev. Res.}\ }\textbf {\bibinfo {volume} {2}},\ \bibinfo {pages} {032044} (\bibinfo {year} {2020})}\BibitemShut {NoStop}%
\bibitem [{\citenamefont {Burch}\ \emph {et~al.}(2018)\citenamefont {Burch}, \citenamefont {Mandrus},\ and\ \citenamefont {Park}}]{Burch2018}%
  \BibitemOpen
  \bibfield  {author} {\bibinfo {author} {\bibfnamefont {K.~S.}\ \bibnamefont {Burch}}, \bibinfo {author} {\bibfnamefont {D.}~\bibnamefont {Mandrus}}, \ and\ \bibinfo {author} {\bibfnamefont {J.-G.}\ \bibnamefont {Park}},\ }\href {\doibase 10.1038/s41586-018-0631-z} {\bibfield  {journal} {\bibinfo  {journal} {Nature}\ }\textbf {\bibinfo {volume} {563}},\ \bibinfo {pages} {47} (\bibinfo {year} {2018})}\BibitemShut {NoStop}%
\bibitem [{\citenamefont {Gibertini}\ \emph {et~al.}(2019)\citenamefont {Gibertini}, \citenamefont {Koperski}, \citenamefont {Morpurgo},\ and\ \citenamefont {Novoselov}}]{Gibertini2019}%
  \BibitemOpen
  \bibfield  {author} {\bibinfo {author} {\bibfnamefont {M.}~\bibnamefont {Gibertini}}, \bibinfo {author} {\bibfnamefont {M.}~\bibnamefont {Koperski}}, \bibinfo {author} {\bibfnamefont {A.~F.}\ \bibnamefont {Morpurgo}}, \ and\ \bibinfo {author} {\bibfnamefont {K.~S.}\ \bibnamefont {Novoselov}},\ }\href {\doibase 10.1038/s41565-019-0438-6} {\bibfield  {journal} {\bibinfo  {journal} {Nature Nanotechnology}\ }\textbf {\bibinfo {volume} {14}},\ \bibinfo {pages} {408} (\bibinfo {year} {2019})}\BibitemShut {NoStop}%
\bibitem [{\citenamefont {Geim}\ and\ \citenamefont {Grigorieva}(2013)}]{Geim2013}%
  \BibitemOpen
  \bibfield  {author} {\bibinfo {author} {\bibfnamefont {A.~K.}\ \bibnamefont {Geim}}\ and\ \bibinfo {author} {\bibfnamefont {I.~V.}\ \bibnamefont {Grigorieva}},\ }\href {\doibase 10.1038/nature12385} {\bibfield  {journal} {\bibinfo  {journal} {Nature}\ }\textbf {\bibinfo {volume} {499}},\ \bibinfo {pages} {419} (\bibinfo {year} {2013})}\BibitemShut {NoStop}%
\bibitem [{\citenamefont {Murzaliev}\ \emph {et~al.}(2024)\citenamefont {Murzaliev}, \citenamefont {Titov},\ and\ \citenamefont {Katsnelson}}]{Supp}%
  \BibitemOpen
  \bibfield  {author} {\bibinfo {author} {\bibfnamefont {B.}~\bibnamefont {Murzaliev}}, \bibinfo {author} {\bibfnamefont {M.}~\bibnamefont {Titov}}, \ and\ \bibinfo {author} {\bibfnamefont {M.}~\bibnamefont {Katsnelson}},\ }\href@noop {} {\enquote {\bibinfo {title} {Supplementary material},}\ }\bibinfo {howpublished} {Attached to the submission of the article to Phys. Rev. Lett.} (\bibinfo {year} {2024})\BibitemShut {NoStop}%
\bibitem [{\citenamefont {Canali}\ and\ \citenamefont {Girvin}(1992)}]{Girvin}%
  \BibitemOpen
  \bibfield  {author} {\bibinfo {author} {\bibfnamefont {C.~M.}\ \bibnamefont {Canali}}\ and\ \bibinfo {author} {\bibfnamefont {S.~M.}\ \bibnamefont {Girvin}},\ }\href@noop {} {\bibfield  {journal} {\bibinfo  {journal} {Phys. Rev. B}\ }\textbf {\bibinfo {volume} {45}},\ \bibinfo {pages} {7127} (\bibinfo {year} {1992})}\BibitemShut {NoStop}%
\end{thebibliography}%

\supplementarystart

\centerline{\bfseries\large ONLINE SUPPLEMENTAL MATERIAL}
\vspace{6pt}
\centerline{\bfseries\large Optical detection of 4-spin chiral interaction in a 2D honeycomb ferrimagnet}
\vspace{6pt}
\centerline{B.\,Murzaliev, M.\,I.\,Katsnelson,  M.\,Titov}
\begin{quote}
This supplementary material provides detailed theoretical derivations and supporting analyses for the main text, 'Optical detection of 4-spin chiral interaction in a 2D honeycomb ferrimagnet.' We focus on the effective ferrimagnetic model of Fe$_{3}$GeTe$_{2}$, detailing the spin Hamiltonian (Sec. I), magnon spectra (Sec. II), interaction terms (Sec. III), and collision integrals (Sec. IV) that underpin the proposed optical detection of the 4-spin chiral interaction.
\end{quote}

\section{Hamiltonian}
This section derives the bosonized Hamiltonian from the spin model in Eq. (1) of the main text. We use the Holstein-Primakoff transformation to express the Hamiltonian in terms of magnon operators ($a^{\dagger}_{\bs{k}}$, $a_{\bs{k}}$, $b^{\dagger}_{\bs{k}}$, $b_{\bs{k}} $). 
Here we present in more detail the mathematical derivations of the results discussed in the main text. Specifically, we derive the spin Hamiltonian interaction in a ferrimagnet in terms of the operators \(a^{\dagger}_{\bs{k}}, a_{\bs{k}}, b^{\dagger}_{\bs{k}}, b_{\bs{k}}\).

\begin{equation}
{\cal H}= -J\sum_{\langle i j \rangle}\boldsymbol{S}_{i}\boldsymbol{S}_{j} - \sum_{i}HS^{z}_{i} -h(t)\sum_{i}S_{i}^{x}+{\cal H}_{{\textrm{4S}}},
\label{eq:HamSpin}
\end{equation}

\begin{equation}
\begin{aligned}
{\cal H}_{\textrm{4S}}=&\frac{A_{\textrm{D}}}{a^4}\sum_{\bs{R}\in O}\lt[\frac{1}{(S^{\textrm{A}})^{3}S^{\textrm{B}}}\prod_{\alpha=1}^{3}\bb{\delta}_{\alpha}\cdot\bs{S}_{\bs{R}+\bb{\delta}_{\alpha}}^{\textrm{A}}\sum_{\beta}\bb{\delta}_{\beta}\cdot\bs{S}_{\bs{R}+\bb{\delta}_{\beta}}^{\textrm{B}}\rt.-\lt.\frac{1}{S^{\textrm{A}}(S^{\textrm{B}})^{3}}\prod_{\beta=1}^{3}\bb{\delta}_{\beta}\cdot\bs{S}_{\bs{R}+\bb{\delta}_{\beta}}^{\textrm{B}}\sum_{\alpha}\bs{\delta}_{\alpha}\cdot\bs{S}_{\bs{R}+\bs{\delta}_{\alpha}}^{\textrm{A}}\rt],
\end{aligned}
\label{eq:Ham4S}
\end{equation}
The first term in the Hamiltonian (\ref{eq:HamSpin}) represents the Heisenberg exchange interaction between spins of different sublattices with exchange constant $J$. The second term is the Zeeman term, which accounts for the application of a permanent magnetic field $H$ to the spins in both sublattices. The third term describes the effect of an applied laser pulse on sublattice A, represented as a transverse effective alternating magnetic field $h(t)$. We assume that the amplitude of this field, $h_{0}$, is much smaller than both $J$ and $H$ $h_{0} \ll J, H$. A new term in the Hamiltonian, ${\cal H}_{{\textrm{4S}}}$, represents the 4-spin interaction with energy constant $A_{\textrm{D}}$. The sum \(\langle ij \rangle \) runs over bonds in a honeycomb lattice with lattice spacing \(a\). Sites with coordinates \(i\) belong to sublattice A, while sites with coordinates \(j\) belong to sublattice B. We define the lattice with A and B sublattices such that the three vectors from an A site to its three nearest B neighbors are given by:

\begin{equation}
\boldsymbol{\delta}_{1} = a\begin{pmatrix}1 \\ 0\end{pmatrix}, \quad \boldsymbol{\delta}_{2} = \frac{a}{2}\begin{pmatrix}-1 \\ \sqrt{3}\end{pmatrix}, \quad \boldsymbol{\delta}_{3} = \frac{a}{2}\begin{pmatrix}-1 \\ -\sqrt{3}\end{pmatrix}.
\end{equation}

In order to apply standard many-body techniques for boson systems, let us transform the spin Hamiltonian given in Eq. (\ref{eq:HamSpin}) into a boson Hamiltonian using the Holstein-Primakoff transformation. For sublattice A:

\begin{equation}
\begin{aligned}
S_{\textit{i}}^{zA} &= S^{\textrm{A}} - a_{\textit{i}}^{\dagger}a_{\textit{i}}, \\
S_{\textit{i}}^{+A} &= \sqrt{2S^{\textrm{A}}-a_{\textit{i}}^{\dagger}a_{\textit{i}}}\,a_{\textit{i}}, \\
S_{\textit{i}}^{-A} &= a_{\textit{i}}^{\dagger}\sqrt{2S^{\textrm{A}}-a_{\textit{i}}^{\dagger}a_{\textit{i}}}. \\
\label{eq:HPA}
\end{aligned}
\end{equation}

For sublattice B:

\begin{equation}
\begin{aligned}
S_{i}^{zB} &= S^{\textrm{B}} - b_{\textit{i}}^{\dagger}b_{\textit{i}}, \\
S_{i}^{+B} &= \sqrt{2S^{\textrm{B}}-b_{\textit{i}}^{\dagger}b_{\textit{i}}}\,b_{\textit{i}}, \\
S_{i}^{-B} &= b_{\textit{i}}^{\dagger}\sqrt{2S^{\textrm{B}}-b_{\textit{i}}^{\dagger}b_{\textit{i}}}. \\
\label{eq:HPB}
\end{aligned}
\end{equation}

This representation of the spin operators obeys the spin commutation relations. By using Eqs. (\ref{eq:HPA}) and (\ref{eq:HPB}) in Eq. (\ref{eq:HamSpin}), the spin Hamiltonian becomes the following Hermitian boson Hamiltonian. We represent the total Hamiltonian in the form of several terms:

\begin{equation}
{\cal H}=E_{0}+{\cal H}_{0}+{\cal H}_{\textrm{ex}}+ {\cal H}_{{\textrm{4S}}},
\end{equation}

The first term is the energy of the ground state (vacuum), where $z = 3)$ for honeycomb lattice, which is the coordination number:

\begin{equation}
E_{0}=NJzS^{\textrm{A}}S^{\textrm{B}},
\end{equation}

The second term is the Hamiltonian of free magnons with applied magnetic fields, including the effective magnetic field from the optical pulse per sublattice:

\begin{equation}
\begin{aligned}
{\cal{H}}_0 = & - J \sum_{\langle ij \rangle} \sqrt{S^{\textrm{A}} S^{\textrm{B}}} \left( a_{\textrm{i}}^{\dagger} b_{\textrm{j}} + a_{\textrm{i}} b_{\textrm{j}}^{\dagger} \right) + (J S^{\textrm{A}} + H) \sum_i b_{\textrm{i}}^{\dagger} b_{\textrm{i}} + (J S^{\textrm{B}} + H) \sum_i a_{\textrm{i}}^{\dagger} a_{\textrm{i}} \\
& - h(t) \left( \sum_i \sqrt{\frac{S^{\textrm{A}}}{2}} \left( a_{\textrm{i}} + a_{\textrm{i}}^{\dagger} \right) + \sqrt{\frac{S^{\textrm{B}}}{2}} \left( b_{\textrm{i}} + b_{\textrm{i}}^{\dagger} \right) \right)
\end{aligned}
\end{equation}

The Hamiltonian responsible for the exchange interaction between magnons of distinct types is:

\begin{equation}
\begin{aligned}
{\cal H}_{\textrm{ex}}&=\frac{J}{4}\sum_{\langle ij\rangle}\Bigg(\sqrt{\frac{S^{\textrm{B}}}{S^{\textrm{A}}}}(a_{\textit{i}}^{\dagger}a_{\textit{i}}^{\dagger}a_{\textit{i}}b_{j}+a_{\textit{i}}^{\dagger}a_{\textit{i}}a_{\textit{i}}b_{j}^{\dagger}) + \sqrt{\frac{S^{\textrm{A}}}{S^{\textrm{B}}}}(a_{\textit{i}}^{\dagger}b_{j}^{\dagger}b_{j}b_{j}+a_{\textit{i}}b_{j}^{\dagger}b_{j}^{\dagger}b_{j})-4a_{\textit{i}}^{\dagger}a_{\textit{i}}b_{j}^{\dagger}b_{j}\Bigg).
\end{aligned}
\label{eq:Hex}
\end{equation}

\section{Spectrum of magnons}
Let us find a spectrum for magnons.  To do that we need to introduce Fourier transformation for the boson operators.

\begin{equation}
\begin{aligned}
a_{\textit{i}}=\frac{1}{\sqrt{N}}\sum_{\bs{k}}e^{-i\boldsymbol{kr}_{i}}a_{\bs{k}},\\
\ensuremath{b_{j}=\frac{1}{\sqrt{N}}\sum_{\bs{k}}e^{-i\boldsymbol{kr}_{j}}b_{\bs{k}}},
\end{aligned}
\end{equation}
where each summation is restricted to the relevant sublattice and
$\bs{k}$ runs over the Brillouin zone, let us rewrite
the quadratic part of the Hamiltonian is given by
\begin{equation}
\begin{aligned}
{\cal H}_{0} &= \sum_{\bs{k}} \left[ (JS^{\textrm{A}} + H) b_{\bs{k}}^{\dagger} b_{\bs{k}} + (JS^{\textrm{B}} + H) a_{\bs{k}}^{\dagger} a_{\bs{k}} - J \sqrt{S^{\textrm{A}} S^{\textrm{B}}} \left( \gamma_{\bs{k}}^{*} a_{\bs{k}} b_{\bs{k}}^{\dagger} + \gamma_{\bs{k}} a_{\bs{k}}^{\dagger} b_{\bs{k}} \right) \right] \\ 
& \quad - h(t) \left( \sqrt{\frac{S^{\textrm{A}}}{2}} \left( a_{\bs{k}=0} + a^{\dagger}_{\bs{k}=0} \right) + \sqrt{\frac{S^{\textrm{B}}}{2}} \left( b_{\bs{k}=0} + b^{\dagger}_{\bs{k}=0} \right) \right),
\end{aligned}
\label{eq:H0}
\end{equation}
with 
\begin{equation}
\gamma_{\bs{k}}=\frac{1}{z}\sum_{\boldsymbol{\delta}}e^{i\boldsymbol{k\delta}}=(2\exp(ik_{x}a/2)\cos(k_{y}a\sqrt{3}/2)+\exp(-ik_{x}a))/3,   
\end{equation}
in the long-wave limit $\gamma_{\bs{k}}=1-(ka)^2/4$. We now proceed to diagonalize (\ref{eq:H0}) using Bogoliubov transformation 
\begin{equation}
\begin{aligned}
a_{\bs{k}}=u_{\bs{k}}\alpha_{\bs{k}}+v_{\bs{k}}\beta_{\bs{k}},
\\
a_{\bs{k}}^{\dagger}=u_{\bs{k}}\alpha_{\bs{k}}^{\dagger}+v_{\bs{k}}\beta_{\bs{k}}^{\dagger}, \\
b_{\bs{k}}=u_{\bs{k}}\beta_{\bs{k}}+v_{\bs{k}}\alpha_{\bs{k}}, \\
b_{\bs{k}}^{\dagger}=u_{\bs{k}}\beta_{\bs{k}}^{\dagger}+v_{\bs{k}}\alpha_{\bs{k}}^{\dagger}.
\end{aligned}
\end{equation}
Let us consider diagonalization of the Hamiltonian ${\cal H}_{0}$, it is the case without the effective magnetic field $h(t) = 0$.

We obtain  the Hamiltonian
\begin{equation}
\begin{aligned}
{\cal H}_{0}=&\sum_{\bs{k}}\lt[\omega^{\alpha}_\bs{k}\alpha_{\bs{k}}^{\dagger}\alpha_{\bs{k}}+\omega^{\beta}_{\bs{k}}\beta_{\bs{k}}^{\dagger}\beta_{\bs{k}}\rt] -h(t)\lt.\lt[A(\alpha_{\bs{k}}+\alpha_{\bs{k}}^{\dagger})+
B(\beta_{\bs{k}}+\beta_{\bs{k}}^{\dagger})\rt]\rt|_{\bs{k}=0},
\end{aligned}
\end{equation}
with the spectrum for $\alpha$-magnons 
\begin{equation}
\omega_{\bs{k}}^{\alpha}=\frac{1}{2} \left( J (-S^{\textrm{A}} + S^{\textrm{B}}) + \sqrt{(2 H + J (S^{\textrm{A}} + S^{\textrm{B}}))^2 - 4 J^2 S^{\textrm{A}} S^{\textrm{B}} \gamma_{\bs{k}}^2} \right),
\end{equation}
and the spectrum for $\beta$-magnons
\begin{equation}
\omega_{\bs{k}}^{\beta}=\frac{1}{2} \left( J (S^{\textrm{A}} - S^{\textrm{B}}) + \sqrt{(2 H + J (S^{\textrm{A}} + S^{\textrm{B}}))^2 - 4 J^2 S^{\textrm{A}} S^{\textrm{B}} \gamma_{\bs{k}}^2} \right),
\end{equation}
coefficients $A$ and $B$ are
\be
A = \sqrt{\frac{S^{\textrm{A}}}{2}} u_{\bs{k}=0} + \sqrt{\frac{S^{\textrm{B}}}{2}} v_{\bs{k}=0}, \quad B = \sqrt{\frac{S^{\textrm{A}}}{2}} v_{\bs{k}=0} + \sqrt{\frac{S^{\textrm{B}}}{2}} u_{\bs{k}=0}
\e
We can see that spectrum for honeycomb lattice without a magnetic field $H$ for one sort of magnon is gapped and for another sort of magnon is gapless, the gap equals $\Delta =J|S^{\textrm{A}}-S^{\textrm{B}}|$ . Both branches exhibit a quadratic dependence  for small $\bs{k}$, in contrast to the antiferromagnetic case, where the magnon spectrum displays a linear dependence.The Bogoliubov coefficients, which characterize the transformation between the original magnon operators and the quasiparticle operators, are 
\begin{equation}
u_{\bs{k}} = \frac{1}{\sqrt{2}}  \sqrt{\frac{2 H + J (S^{\textrm{A}} + S^{\textrm{B}}) + \sqrt{-4 J^2 S^{\textrm{A}}S^{\textrm{B}}\gamma_{\bs{k}}^2 + (2 H + J (S^{\textrm{A}} + S^{\textrm{B}}))^2}}{\sqrt{-4 J^2 S^{\textrm{A}}S^{\textrm{B}} \gamma_{\bs{k}}^2+ (2 H + J (S^{\textrm{A}} + S^{\textrm{B}}))^2}}} ,
\end{equation}

\begin{equation}
v_{\bs{k}} = \left( \frac{\sqrt{2} J \sqrt{S^{\textrm{A}}S^{\textrm{B}}} \gamma_{\bs{k}}}{\sqrt{-4 J^2 S^{\textrm{A}}S^{\textrm{B}} \gamma_{\bs{k}}^2+ (2 H + J (S^{\textrm{A}} + S^{\textrm{B}}))^2} \sqrt{\frac{2 H + J (S^{\textrm{A}} + S^{\textrm{B}}) + \sqrt{-4 J^2 S^{\textrm{A}}S^{\textrm{B}} \gamma_{\bs{k}}^2+ (2 H + J (S^{\textrm{A}} + S^{\textrm{B}}))^2}}{\sqrt{-4 J^2 S^{\textrm{A}}S^{\textrm{B}} \gamma_{\bs{k}}^2+ (2 H + J (S^{\textrm{A}} + S^{\textrm{B}}))^2}}}} \right).
\end{equation}

These coefficients are crucial for understanding the quasiparticle excitations and the resulting magnon dynamics in the system. It is important to note that the coefficients $u_{\bs{k}}$ and $v_{\bs{k}}$ exhibit a divergence in the case of antiferromagnetism when $S^{\textrm{A}} = S^{\textrm{B}}$.  The reason for this is that Taylor's expansion is invalid in this context. Furthermore, for the correct expansion in the case of antiferromagnetism, the coefficient $u_{\bs{k}}$ exhibits a divergence since $u_{\bs{k}}\propto 1/k$ \cite{Girvin} The key distinguishing feature between a ferrimagnet with antiparallel spins and an antiferromagnet is the dependence of the Bogoliubov coefficients on the magnetic field $H$.

\section{Interaction}
Let us consider Hamiltonian with interaction in Fourier space. We have two terms - exchange and four-spin interactions.

\begin{equation}
\begin{aligned}
&{\cal H}_{\text{ex}}=\frac{J}{N}\sum_{\bs{k}_{1}\bs{k}_{2}\bs{k}_{3}\bs{k}_{4}}\delta(\bs{k}_{1}+\bs{k}_{2}-\bs{k}_{3}-\bs{k}_{4})\Big[4a_{\bs{k}_{1}}^{\dagger}b_{\bs{k}_{2}}^{\dagger}a_{\bs{k}_{3}}b_{\bs{k}_{4}}\gamma_{\bs{k}_{2}-\bs{k}_{4}}-1/\eta(\gamma_{\bs{k}_{2}}a_{\bs{k}_{1}}^{\dagger}b_{\bs{k}_{2}}^{\dagger}a_{\bs{k}_{3}}a_{\bs{k}_{4}}\\
&+\gamma_{-\bs{k}_{4}}a_{\bs{k}_{1}}^{\dagger}a_{\bs{k}_{2}}^{\dagger}a_{\bs{k}_{3}}b_{\bs{k}_{4}})-\eta(\gamma_{\bs{k}_{2}-\bs{k}_{3}-\bs{k}_{4}}a_{\bs{k}_{1}}^{\dagger}b_{\bs{k}_{2}}^{\dagger}b_{\bs{k}_{3}}b_{\bs{k}_{4}}+\gamma_{\bs{k}_{1}+\bs{k}_{2}-\bs{k}_{4}}b_{\bs{k}_{1}}^{\dagger}b_{\bs{k}_{2}}^{\dagger}a_{\bs{k}_{3}}b_{\bs{k}_{4}})\Big].
\end{aligned}
\end{equation}
The key distinction from magnon-magnon interaction in antiferromagnets lies in the presence of the factor $\eta = \sqrt{\frac{S^{\textrm{B}}}{S^{\textrm{A}}}}$ in the second and third terms, respectively. Comparing with the Hamiltonian for an antiferromagnet from [reference], it can be concluded that the matrix elements have the same form with the replacement of parameters.  To rewrite the Hamiltonian in terms of operators $\alpha^{\dagger}_{\bs{k}}$, $\alpha_{\bs{k}}$, $\beta^{\dagger}_{\bs{k}}$, $\beta_{\bs{k}}$ we use Bogoliubov transformation

In terms of magnons after Bogoliubov transformation we have Hamiltonian of interaction
\begin{equation}
\begin{aligned}
&{\cal H}_{\text{ex}} =  -\frac{Jz}{4N}\sum_{\bs{k}_{1}\bs{k}_{2}\bs{k}_{3}\bs{k}_{4}}\delta(\bs{k}_{1}+\bs{k}_{2}-\bs{k}_{3}-\bs{k}_{4})u_{\bs{k}_{1}}u_{\bs{k}_{2}}u_{\bs{k}_{3}}u_{\bs{k}_{4}}(V_{\bs{k}_{1}\bs{k}_{2}\bs{k}_{3}\bs{k}_{4}}^{(1)}\alpha_{\bs{k}_{1}}^{\dagger}\alpha_{\bs{k}_{2}}^{\dagger}\alpha_{\bs{k}_{3}}\alpha_{\bs{k}_{4}}+V_{\bs{k}_{1}\bs{k}_{2}\bs{k}_{3}\bs{k}_{4}}^{(2)}\alpha_{\bs{k}_{1}}^{\dagger}\beta_{\bs{k}_{2}}^{\dagger}\alpha_{\bs{k}_{3}}\alpha_{\bs{k}_{4}}\\
&+V_{\bs{k}_{1}\bs{k}_{2}\bs{k}_{3}\bs{k}_{4}}^{(3)}\alpha_{\bs{k}_{1}}^{\dagger}\alpha_{\bs{k}_{2}}^{\dagger}\alpha_{\bs{k}_{3}}\beta_{\bs{k}_{4}}+ V_{\bs{k}_{1}\bs{k}_{2}\bs{k}_{3}\bs{k}_{4}}^{(4)}\alpha_{\bs{k}_{1}}^{\dagger}\beta_{\bs{k}_{2}}^{\dagger}\alpha_{\bs{k}_{3}}\beta_{\bs{k}_{4}} +V_{\bs{k}_{1}\bs{k}_{2}\bs{k}_{3}\bs{k}_{4}}^{(5)}\beta_{\bs{k}_{1}}^{\dagger}\beta_{\bs{k}_{2}}^{\dagger}\alpha_{\bs{k}_{3}}\beta_{\bs{k}_{4}} + 2V_{\bs{k}_{1}\bs{k}_{2}\bs{k}_{3}\bs{k}_{4}}^{(6)}\alpha_{\bs{k}_{1}}^{\dagger}\beta_{\bs{k}_{2}}^{\dagger}\beta_{\bs{k}_{3}}\beta_{\bs{k}_{4}} \\
&  + V_{\bs{k}_{1}\bs{k}_{2}\bs{k}_{3}\bs{k}_{4}}^{(7)}\alpha_{\bs{k}_{1}}^{\dagger}\alpha_{\bs{k}_{2}}^{\dagger}\beta_{\bs{k}_{3}}\beta_{\bs{k}_{4}}+ V_{\bs{k}_{1}\bs{k}_{2}\bs{k}_{3}\bs{k}_{4}}^{(8)}\beta_{\bs{k}_{1}}^{\dagger}\beta_{\bs{k}_{2}}^{\dagger}\alpha_{\bs{k}_{3}}\alpha_{\bs{k}_{4}} + V_{\bs{k}_{1}\bs{k}_{2}\bs{k}_{3}\bs{k}_{4}}^{(9)}\beta_{\bs{k}_{1}}^{\dagger}\beta_{\bs{k}_{2}}^{\dagger}\beta_{\bs{k}_{3}}\beta_{\bs{k}_{4}}).
\end{aligned}
\end{equation}

with matrix elements ($x_{\bs{k}}= -v_{\bs{k}}/u_{\bs{k}}$ )

\begin{equation}
\begin{aligned}
V_{\bs{k}_{1}\bs{k}_{2}\bs{k}_{3}\bs{k}_{4}}^{(1)}=(\gamma_{\bs{k}_{1}-\bs{k}_{4}}x_{\bs{k}_{1}}x_{\bs{k}_{4}}+\gamma_{\bs{k}_{1}-\bs{k}_{3}}x_{\bs{k}_{1}}x_{\bs{k}_{3}}+\gamma_{\bs{k}_{2}-\bs{k}_{4}}x_{\bs{k}_{2}}x_{\bs{k}_{4}}+\gamma_{\bs{k}_{2}-\bs{k}_{3}}x_{\bs{k}_{2}}x_{\bs{k}_{3}}-\\
\eta(\gamma_{\bs{k}_{1}}x_{\bs{k}_{2}}x_{\bs{k}_{3}}x_{\bs{k}_{4}}+\gamma_{\bs{k}_{2}}x_{\bs{k}_{1}}x_{\bs{k}_{3}}x_{\bs{k}_{4}})-1/\eta(\gamma_{\bs{k}_{2}}x_{\bs{k}_{2}}+\gamma_{\bs{k}_{1}}x_{\bs{k}_{1}})),
\end{aligned}
\end{equation}

\begin{equation}
\begin{aligned}
V_{\bs{k}_{1}\bs{k}_{2}\bs{k}_{3}\bs{k}_{4}}^{(2)}=2(-\gamma_{\bs{k}_{1}-\bs{k}_{4}}x_{\bs{k}_{1}}x_{\bs{k}_{2}}x_{\bs{k}_{4}}-\gamma_{\bs{k}_{1}-\bs{k}_{3}}x_{\bs{k}_{1}}x_{\bs{k}_{2}}x_{\bs{k}_{3}}-x_{\bs{k}_{4}}\gamma_{\bs{k}_{2}-\bs{k}_{4}}-x_{\bs{k}_{3}}\gamma_{\bs{k}_{2}-\bs{k}_{3}}+\\
\eta(\gamma_{\bs{k}_{2}}x_{\bs{k}_{1}}x_{\bs{k}_{2}}x_{\bs{k}_{3}}x_{\bs{k}_{4}}+\gamma_{\bs{k}_{1}}x_{\bs{k}_{3}}x_{\bs{k}_{4}})+1/\eta(\gamma x_{\bs{k}_{1}}x_{\bs{k}_{2}}+\gamma_{\bs{k}_{2}}))
\end{aligned}
\end{equation}

\begin{equation}
\begin{aligned}
V_{\bs{k}_{1}\bs{k}_{2}\bs{k}_{3}\bs{k}_{4}}^{(3)}=2(-\gamma_{\bs{k}_{2}-\bs{k}_{4}}x_{\bs{k}_{2}}-\gamma_{\bs{k}_{1}-\bs{k}_{4}}x_{\bs{k}_{1}}-\gamma_{\bs{k}_{2}-\bs{k}_{4}}x_{\bs{k}_{1}}x_{\bs{k}_{3}}x_{\bs{k}_{4}}-\gamma_{\bs{k}_{2}-\bs{k}_{3}}x_{\bs{k}_{2}}x_{\bs{k}_{3}}x_{\bs{k}_{4}}+\\
\eta(\gamma_{\bs{k}_{1}}x_{\bs{k}_{2}}x_{\bs{k}_{3}}+\gamma_{\bs{k}_{2}}x_{\bs{k}_{1}}x_{\bs{k}_{3}})+1/\eta(\gamma_{\bs{k}_{2}}x_{\bs{k}_{2}}x_{\bs{k}_{4}}+\gamma_{\bs{k}_{1}}x_{\bs{k}_{1}}x_{\bs{k}_{4}})),
\end{aligned}
\end{equation}

\begin{equation}
\begin{aligned}
V_{\bs{k}_{1}\bs{k}_{2}\bs{k}_{3}\bs{k}_{4}}^{(4)}=4(\gamma_{\bs{k}_{2}-\bs{k}_{4}}+\gamma_{\bs{k}_{1}-\bs{k}_{4}}x_{\bs{k}_{1}}x_{\bs{k}_{2}}+\gamma_{\bs{k}_{2}-\bs{k}_{3}}x_{\bs{k}_{3}}x_{\bs{k}_{4}}+\gamma_{\bs{k}_{1}-\bs{k}_{3}}x_{\bs{k}_{1}}x_{\bs{k}_{2}}x_{\bs{k}_{3}}x_{\bs{k}_{4}}-\\
\eta(\gamma_{\bs{k}_{2}}x_{\bs{k}_{1}}x_{\bs{k}_{2}}x_{\bs{k}_{3}}-\gamma_{\bs{k}_{1}}x_{\bs{k}_{3}})-\frac{1}{\eta}(\gamma_{\bs{k}_{1}}x_{\bs{k}_{1}}x_{\bs{k}_{2}}x_{\bs{k}_{4}}-\gamma_{\bs{k}_{2}}x_{\bs{k}_{4}})),
\end{aligned}
\end{equation}

\begin{equation}
\begin{aligned}
V_{\bs{k}_{1}\bs{k}_{2}\bs{k}_{3}\bs{k}_{4}}^{(5)}=(-\gamma_{\bs{k}_{2}-\bs{k}_{4}}x_{\bs{k}_{1}}-\gamma_{\bs{k}_{1}-\bs{k}_{4}}x_{\bs{k}_{2}}-\gamma_{\bs{k}_{2}-\bs{k}_{3}}x_{\bs{k}_{1}}x_{\bs{k}_{3}}x_{\bs{k}_{4}}-\gamma_{\bs{k}_{1}-\bs{k}_{3}}x_{\bs{k}_{2}}x_{\bs{k}_{3}}x_{\bs{k}_{4}}+\\
\frac{1}{\eta}(\gamma_{\bs{k}_{1}}x_{\bs{k}_{2}}x_{\bs{k}_{4}}+\gamma_{\bs{k}_{2}}x_{\bs{k}_{1}}x_{\bs{k}_{4}})+\eta(\gamma_{\bs{k}_{2}}x_{\bs{k}_{2}}x_{\bs{k}_{3}}+\gamma_{\bs{k}_{1}}x_{\bs{k}_{1}}x_{\bs{k}_{3}})),
\end{aligned}
\end{equation}

\begin{equation}
\begin{aligned}
V_{\bs{k}_{1}\bs{k}_{2}\bs{k}_{3}\bs{k}_{4}}^{(6)}=2(-\gamma_{\bs{k}_{2}-\bs{k}_{4}}x_{\bs{k}_{1}}x_{\bs{k}_{2}}x_{\bs{k}_{4}}-\gamma_{\bs{k}_{2}-\bs{k}_{3}}x_{\bs{k}_{1}}x_{\bs{k}_{2}}x_{\bs{k}_{3}}-x_{\bs{k}_{4}}\gamma_{\bs{k}_{1}-\bs{k}_{4}}-x_{\bs{k}_{3}}\gamma_{\bs{k}_{1}-\bs{k}_{3}}+\\
\frac{1}{\eta}(\gamma_{\bs{k}_{2}}x_{\bs{k}_{1}}x_{\bs{k}_{2}}x_{\bs{k}_{3}}x_{\bs{k}_{4}}+\gamma_{\bs{k}_{1}}x_{\bs{k}_{3}}x_{\bs{k}_{4}})+\eta(\gamma_{\bs{k}_{1}}x_{\bs{k}_{1}}x_{\bs{k}_{2}}+\gamma_{\bs{k}_{2}})),
\end{aligned}
\end{equation}

\begin{equation}
\begin{aligned}
V_{\bs{k}_{1}\bs{k}_{2}\bs{k}_{3}\bs{k}_{4}}^{(7)}=(x_{\bs{k}_{1}}x_{\bs{k}_{3}}\gamma_{\bs{k}_{1}-\bs{k}_{4}}+x_{\bs{k}_{1}}x_{\bs{k}_{4}}\gamma_{\bs{k}_{1}-\bs{k}_{3}}+x_{\bs{k}_{2}}x_{\bs{k}_{3}}\gamma_{\bs{k}_{2}-\bs{k}_{4}}+x_{\bs{k}_{2}}x_{\bs{k}_{4}}\gamma_{\bs{k}_{2}-\bs{k}_{3}}-\\
\eta(\gamma_{\bs{k}_{2}}x_{\bs{k}_{1}}-\gamma_{\bs{k}_{1}}x_{\bs{k}_{2}})-\frac{1}{\eta}(\gamma_{\bs{k}_{2}}x_{\bs{k}_{2}}x_{\bs{k}_{3}}x_{\bs{k}_{4}}-\gamma_{\bs{k}_{1}}x_{\bs{k}_{1}}x_{\bs{k}_{3}}x_{\bs{k}_{4}})),
\end{aligned}
\end{equation}

\begin{equation}
\begin{aligned}
V_{\bs{k}_{1}\bs{k}_{2}\bs{k}_{3}\bs{k}_{4}}^{(8)}=(x_{\bs{k}_{1}}x_{\bs{k}_{3}}\gamma_{\bs{k}_{2}-\bs{k}_{3}}+x_{\bs{k}_{1}}x_{\bs{k}_{4}}\gamma_{\bs{k}_{2}-\bs{k}_{4}}+x_{\bs{k}_{2}}x_{\bs{k}_{3}}\gamma_{\bs{k}_{1}-\bs{k}_{3}}+x_{\bs{k}_{2}}x_{\bs{k}_{4}}\gamma_{\bs{k}_{1}-\bs{k}_{4}}-\\
\eta(\gamma_{\bs{k}_{2}}x_{\bs{k}_{2}}x_{\bs{k}_{3}}x_{\bs{k}_{4}}-\gamma_{\bs{k}_{1}}x_{\bs{k}_{1}}x_{\bs{k}_{3}}x_{\bs{k}_{4}})-\frac{1}{\eta}(\gamma_{\bs{k}_{2}}x_{\bs{k}_{1}}-\gamma_{\bs{k}_{1}}x_{\bs{k}_{2}})),
\end{aligned}
\end{equation}

\begin{equation}
\begin{aligned}
V_{\bs{k}_{1}\bs{k}_{2}\bs{k}_{3}\bs{k}_{4}}^{(9)}=(\gamma_{\bs{k}_{2}-\bs{k}_{3}}x_{\bs{k}_{1}}x_{\bs{k}_{4}}+\gamma_{\bs{k}_{2}-\bs{k}_{4}}x_{\bs{k}_{1}}x_{\bs{k}_{3}}+\gamma_{\bs{k}_{1}-\bs{k}_{3}}x_{\bs{k}_{2}}x_{\bs{k}_{4}}+\gamma_{\bs{k}_{1}-\bs{k}_{4}}x_{\bs{k}_{2}}x_{\bs{k}_{4}}-\\
\frac{1}{\eta}(\gamma_{\bs{k}_{1}}x_{\bs{k}_{2}}x_{\bs{k}_{3}}x_{\bs{k}_{4}}-\gamma_{\bs{k}_{2}}x_{\bs{k}_{1}}x_{\bs{k}_{3}}x_{\bs{k}_{4}})-\eta(\gamma_{\bs{k}_{2}}x_{\bs{k}_{2}}-\gamma_{\bs{k}_{1}}x_{\bs{k}_{1}})).
\end{aligned}
\end{equation}
These matrix  have a symmetry for a  case $\eta=1$ : $V_{\bs{k}_{1}\bs{k}_{2}\bs{k}_{3}\bs{k}_{4}}^{(1)}=V_{\bs{k}_{1}\bs{k}_{2}\bs{k}_{3}\bs{k}_{4}}^{(9)}$, $V_{\bs{k}_{1}\bs{k}_{2}\bs{k}_{3}\bs{k}_{4}}^{(2)}=V_{\bs{k}_{2}\bs{k}_{1}\bs{k}_{3}\bs{k}_{4}}^{(6)}$, $V_{\bs{k}_{1}\bs{k}_{2}\bs{k}_{3}\bs{k}_{4}}^{(3)}=V_{\bs{k}_{1}\bs{k}_{2}\bs{k}_{4}\bs{k}_{3}}^{(5)}$ , $V_{\bs{k}_{1}\bs{k}_{2}\bs{k}_{4}\bs{k}_{3}}^{(7)}=V_{\bs{k}_{1}\bs{k}_{2}\bs{k}_{3}\bs{k}_{4}}^{(8)}$. For a ferrimagnetic case $0<\eta<1$ we have violation the symmetry in these matrix elements. These matrix elements look like in Halperin for the antiferromagnetic case $\eta = 1$ and $\beta_{\bs{k}} \Longrightarrow \beta_{-\bs{k}}^{\dagger}$, $\beta_{\bs{k}}^\dagger \Longrightarrow \beta_{-\bs{k}}$.

Using Holstein-Primakoff representation we can write the Hamiltonian of four-spin interaction in magnon representation

\begin{equation}
\begin{aligned}
{\cal H}_{{\textrm{4S}}} = \frac{A_{D}}{4(S^{\textrm{A}}S^{\textrm{B}})^{3/2}}\sum_{\boldsymbol{R}}[(a_{\boldsymbol{R}+\boldsymbol{\delta}_{1}}+a_{\boldsymbol{R}+\boldsymbol{\delta}_{1}}^{\dagger})(e^{i\pi/3}a_{\boldsymbol{R}+\boldsymbol{\delta}_{2}}+e^{-i\pi/3}a_{\boldsymbol{R}+\boldsymbol{\delta}_{2}}^{\dagger})(e^{-i\pi/3}a_{\boldsymbol{R}+\boldsymbol{\delta}_{3}}+e^{i\pi/3}a_{\boldsymbol{R}+\boldsymbol{\delta}_{3}}^{\dagger})\\
\cdot (b_{\boldsymbol{R}+\boldsymbol{\delta}_{1}}-e^{i\pi/3}b_{\boldsymbol{R}+\boldsymbol{\delta}_{2}}-e^{-i\pi/3}b_{\boldsymbol{R}+\boldsymbol{\delta}_{3}}+b_{\boldsymbol{R}+\boldsymbol{\delta}_{1}}^{\dagger}-e^{-i\pi/3}b_{\boldsymbol{R}+\boldsymbol{\delta}_{2}}^{\dagger}-e^{i\pi/3}b_{\boldsymbol{R}+\boldsymbol{\delta}_{3}}^{\dagger})S^{\textrm{B}} - \\
((a_{\boldsymbol{R}+\boldsymbol{\delta}_{1}}-e^{i\pi/3}a_{\boldsymbol{R}+\boldsymbol{\delta}_{2}}-e^{-i\pi/3}a_{\boldsymbol{R}+\boldsymbol{\delta}_{3}}+a_{\boldsymbol{R}+\boldsymbol{\delta}_{1}}^{\dagger}-e^{-i\pi/3}a_{\boldsymbol{R}+\boldsymbol{\delta}_{2}}^{\dagger}-e^{i\pi/3}a_{\boldsymbol{R}+\boldsymbol{\delta}_{3}}^{\dagger})\\
\cdot(b_{\boldsymbol{R}+\boldsymbol{\delta}_{1}}+b_{\boldsymbol{R}+\boldsymbol{\delta}_{1}}^{\dagger})(e^{i\pi/3}b_{\boldsymbol{R}+\boldsymbol{\delta}_{2}}+e^{-i\pi/3}b_{\boldsymbol{R}+\boldsymbol{\delta}_{2}}^{\dagger})(e^{-i\pi/3}b_{\boldsymbol{R}+\boldsymbol{\delta}_{3}}+e^{i\pi/3}b_{\boldsymbol{R}+\boldsymbol{\delta}_{3}}^{\dagger})S^{\textrm{A}}].
\end{aligned}
\end{equation}
Origin of angles $\pi/3$ comes from hexaongal structure of the lattice. 
After Fourier and  Bogoiulubov transformation we leave terms corresponding matrix elements  corresponding three $\alpha$  magnons and one $\beta$ magnon. 
\begin{equation}
\begin{aligned}
&{\cal H}_{{\textrm{4S}}} = \frac{A_{\textrm{D}}}{32(S^{\textrm{A}}S^{\textrm{B}})^{3/2}} \Bigg[\sum_{\bs{k}_{1}\bs{k}_{2}\bs{k}_{3}\bs{k}_{4}}\delta(\bs{k}_{1}+\bs{k}_{2}+\bs{k}_{3}+\bs{k}_{4})\tilde{U}_{\bs{k}_{1},\bs{k}_{2},\bs{k}_{3},\bs{k}_{4}}\alpha_{\bs{k}_{1}}^{\dagger}\alpha_{\bs{k}_{2}}^{\dagger}\alpha_{\bs{k}_{3}}^{\dagger}\beta_{\bs{k}_{4}}] +h.c. \\
\end{aligned}
\end{equation}
Let us denote $\Gamma_{\bs{k}} = (e^{i\bs{k}\boldsymbol{\delta}_{1}}-e^{i\bs{k}\boldsymbol{\delta}_{2}}e^{i\pi/3}-e^{i\bs{k}\boldsymbol{\delta}_{3}}e^{-i\pi/3})$, $U_{\bs{k}_{1},\bs{k}_{2},\bs{k}_{3},\bs{k}_{4}}=e^{i\bs{k}_{1}\boldsymbol{\delta}_{1}}e^{i\bs{k}_{2}\boldsymbol{\delta}_{2}}e^{i\bs{k}_{3}\boldsymbol{\delta}_{3}}\Gamma_{\bs{k}_{4}}$,  and $ \bs{k}_{1}, \bs{k}_{2}, \bs{k}_{3} , \bs{k}_{4}=1,2,3,4 $. The matrix elements are represented in the following way

\begin{equation}
\begin{aligned}
\tilde{U}_{1234} ={} & -S^{\textrm{A}} \big( 
e^{2\pi i/3} v_1 v_2 u_3 u_4 \, U_{-1,-2,4,3}
+ v_1 v_2 v_3 v_4 \, U_{-1,-2,-3,-4} + u_1 v_2 v_3 u_4 \, U_{4,-2,-3,1}
+ e^{-2\pi i/3} v_1 u_2 v_3 u_4 \, U_{-1,4,-3,2}
\big) \\
& + S^{\textrm{B}} \big(
u_1 u_2 u_3 u_4 \, U_{1,2,3,4}
+ v_1 u_2 u_3 v_4 \, U_{-4,2,3,-1} + e^{2\pi i/3} u_1 v_2 u_3 v_4 \, U_{1,-4,3,-2}
+ e^{-2\pi i/3} u_1 u_2 v_3 v_4 \, U_{1,2,-4,-3}
\big)
\end{aligned}
\end{equation}

\section{Collision terms}
Let us express explicitly collision terms for scattering for $\alpha$ and $\beta$ magnons. The collision integral $I^{\textrm{ex}}_{\alpha\alpha}$ consist of two contributions: 1) two $\alpha$-magnons scatter into $\alpha$-magnons and 2) one $\alpha$ and one $\beta$-magnon scatter into another pair of $\alpha$ and $\beta$-magnons.

\begin{equation}
\begin{aligned}
I^{\textrm{ex}}_{\alpha\alpha} ={} & 
2\pi \left( \frac{Jz}{4N} \right)^2 \frac{1}{V_D^2}
\int \frac{d^2\bs{k}_2}{(2\pi)^2} \frac{d^2\bs{k}_3}{(2\pi)^2} \frac{d^2\bs{k}_4}{(2\pi)^2}
\, \delta(\bs{k} + \bs{k}_2 - \bs{k}_3 - \bs{k}_4) \times \bigg\{ \\[4pt]
& \delta(\omega_{\bs{k}}^{\alpha} + \omega_{\bs{k}_2}^{\alpha} - \omega_{\bs{k}_3}^{\alpha} - \omega_{\bs{k}_4}^{\alpha})
\left( V^{(1)}_{\bs{k}, \bs{k}_2, \bs{k}_3, \bs{k}_4} \right)^2
\left[ 
(1 + N_{\bs{k}}^{\alpha})(1 + N_{\bs{k}_2}^{\alpha}) N_{\bs{k}_3}^{\alpha} N_{\bs{k}_4}^{\alpha} 
- N_{\bs{k}}^{\alpha} N_{\bs{k}_2}^{\alpha} (1 + N_{\bs{k}_3}^{\alpha})(1 + N_{\bs{k}_4}^{\alpha})
\right] \\[4pt]
& + \delta(\omega_{\bs{k}}^{\alpha} + \omega_{\bs{k}_2}^{\beta} - \omega_{\bs{k}_3}^{\alpha} - \omega_{\bs{k}_4}^{\beta})
\left( V^{(4)}_{\bs{k}, \bs{k}_2, \bs{k}_3, \bs{k}_4} \right)^2
\left[ 
(1 + N_{\bs{k}}^{\alpha})(1 + N_{\bs{k}_2}^{\beta}) N_{\bs{k}_3}^{\alpha} N_{\bs{k}_4}^{\beta} 
- N_{\bs{k}}^{\alpha} N_{\bs{k}_2}^{\beta} (1 + N_{\bs{k}_3}^{\alpha})(1 + N_{\bs{k}_4}^{\beta})
\right]
\bigg\}
\end{aligned}
\end{equation}

where $V_{D}$ - volume of the unit cell in $k$-space.
Similarly, the collision term $I^{\textrm{ex}}_{\beta\beta}$ consist of : 1) two $\beta$-magnons scatter into other two $\beta$-magnons and 2) one $\beta$ and one $\alpha$-magnons scatter into another pair $\beta$ and $\alpha$-magnons.

\begin{equation}
\begin{aligned}
I^{\textrm{ex}}_{\beta\beta} ={} &
2\pi \left( \frac{Jz}{4N} \right)^2 \frac{1}{V_D^{3/2}}
\int \frac{d^2\bs{k}_2}{(2\pi)^2} \frac{d^2\bs{k}_3}{(2\pi)^2} \frac{d^2\bs{k}_4}{(2\pi)^2}
\, \delta(\bs{k} + \bs{k}_2 - \bs{k}_3 - \bs{k}_4) \times \bigg\{ \\[4pt]
& \delta(\omega_{\bs{k}}^{\beta} + \omega_{\bs{k}_2}^{\beta} - \omega_{\bs{k}_3}^{\beta} - \omega_{\bs{k}_4}^{\beta})
\left( V^{(9)}_{\bs{k}, \bs{k}_2, \bs{k}_3, \bs{k}_4} \right)^2
\left[
(1 + N_{\bs{k}}^{\beta})(1 + N_{\bs{k}_2}^{\beta}) N_{\bs{k}_3}^{\beta} N_{\bs{k}_4}^{\beta} 
- N_{\bs{k}}^{\beta} N_{\bs{k}_2}^{\beta} (1 + N_{\bs{k}_3}^{\beta})(1 + N_{\bs{k}_4}^{\beta})
\right] \\[4pt]
& + \delta(\omega_{\bs{k}}^{\beta} + \omega_{\bs{k}_2}^{\alpha} - \omega_{\bs{k}_3}^{\beta} - \omega_{\bs{k}_4}^{\alpha})
\left( V^{(4)}_{\bs{k}, \bs{k}_2, \bs{k}_3, \bs{k}_4} \right)^2
\left[
(1 + N_{\bs{k}}^{\beta})(1 + N_{\bs{k}_2}^{\alpha}) N_{\bs{k}_3}^{\beta} N_{\bs{k}_4}^{\alpha} 
- N_{\bs{k}}^{\beta} N_{\bs{k}_2}^{\alpha} (1 + N_{\bs{k}_3}^{\beta})(1 + N_{\bs{k}_4}^{\alpha})
\right]
\bigg\}
\end{aligned}
\end{equation}

In contrast to the above integrals, the following terms describe scatteting processes in which  one of the magnons change a sort after scatteting: 1) two $\alpha$-magnons scatter into one $\alpha$-magnon and one $\beta$-magnon, 2) one $\alpha$-magnon and one $\beta$-magnon scatter into two $\alpha$-magnons. These processes are captured by the integral $I^{\textrm{ex}}_{\alpha\beta}$
\begin{equation}
\begin{aligned}
I^{\textrm{ex}}_{\alpha\beta} ={} &
2\pi \left( \frac{Jz}{4N} \right)^2 \frac{1}{V_D^{2}} 
\int \frac{d^2\bs{k}_2}{(2\pi)^2} \frac{d^2\bs{k}_3}{(2\pi)^2} \frac{d^2\bs{k}_4}{(2\pi)^2}
\, \delta(\bs{k} + \bs{k}_2 - \bs{k}_3 - \bs{k}_4) \times \bigg\{ \\[4pt]
& \delta(\omega_{\bs{k}}^{\alpha} + \omega_{\bs{k}_2}^{\alpha} - \omega_{\bs{k}_3}^{\alpha} - \omega_{\bs{k}_4}^{\beta}) 
V^{(2)}_{\bs{k}, \bs{k}_2, \bs{k}_3, \bs{k}_4} 
V^{(3)}_{\bs{k}_4, \bs{k}_3, \bs{k}_2, \bs{k}} 
\left[
(1 + N_{\bs{k}}^{\alpha})(1 + N_{\bs{k}_2}^{\alpha}) N_{\bs{k}_3}^{\alpha} N_{\bs{k}_4}^{\beta} 
- N_{\bs{k}}^{\alpha} N_{\bs{k}_2}^{\alpha} (1 + N_{\bs{k}_3}^{\alpha})(1 + N_{\bs{k}_4}^{\beta})
\right] \\[4pt]
& + \delta(\omega_{\bs{k}}^{\alpha} + \omega_{\bs{k}_2}^{\beta} - \omega_{\bs{k}_3}^{\beta} - \omega_{\bs{k}_4}^{\beta}) 
V^{(6)}_{\bs{k}, \bs{k}_2, \bs{k}_3, \bs{k}_4} 
V^{(5)}_{\bs{k}_4, \bs{k}_3, \bs{k}_2, \bs{k}} 
\left[
(1 + N_{\bs{k}}^{\alpha})(1 + N_{\bs{k}_2}^{\beta}) N_{\bs{k}_3}^{\beta} N_{\bs{k}_4}^{\beta} 
- N_{\bs{k}}^{\alpha} N_{\bs{k}_2}^{\beta} (1 + N_{\bs{k}_3}^{\beta})(1 + N_{\bs{k}_4}^{\beta})
\right]
\bigg\}
\end{aligned}
\end{equation}

The same approach applies for $I^{\textrm{ex}}_{\beta\alpha}$ , which describes: 1) one $\beta$-magnon and $\alpha$-magnon scatter into two $\alpha$-magnons, 2) one $\beta$-magnon and one $\alpha$-magnon scatter into two $\beta$-magnons. These processes are expressed in the integral $I^{\textrm{ex}}_{\beta\alpha}$

\begin{equation}
\begin{aligned}
I^{\textrm{ex}}_{\beta\alpha} ={} &
2\pi \left( \frac{Jz}{4N} \right)^2 \frac{1}{V_D^{2}} 
\int \frac{d^2\bs{k}_2}{(2\pi)^2} \frac{d^2\bs{k}_3}{(2\pi)^2} \frac{d^2\bs{k}_4}{(2\pi)^2}
\, \delta(\bs{k} + \bs{k}_2 - \bs{k}_3 - \bs{k}_4) \times \bigg\{ \\[4pt]
& \delta(\omega_{\bs{k}}^{\beta} + \omega_{\bs{k}_2}^{\alpha} - \omega_{\bs{k}_3}^{\alpha} - \omega_{\bs{k}_4}^{\alpha}) 
V^{(2)}_{\bs{k}, \bs{k}_2, \bs{k}_3, \bs{k}_4} 
V^{(3)}_{\bs{k}_4, \bs{k}_3, \bs{k}_2, \bs{k}} 
\left[
(1 + N_{\bs{k}}^{\beta})(1 + N_{\bs{k}_2}^{\alpha}) N_{\bs{k}_3}^{\alpha} N_{\bs{k}_4}^{\alpha} 
- N_{\bs{k}}^{\beta} N_{\bs{k}_2}^{\alpha} (1 + N_{\bs{k}_3}^{\alpha})(1 + N_{\bs{k}_4}^{\alpha})
\right] \\[4pt]
& + \delta(\omega_{\bs{k}}^{\beta} + \omega_{\bs{k}_2}^{\alpha} - \omega_{\bs{k}_3}^{\beta} - \omega_{\bs{k}_4}^{\beta}) 
V^{(6)}_{\bs{k}, \bs{k}_2, \bs{k}_3, \bs{k}_4} 
V^{(5)}_{\bs{k}_4, \bs{k}_3, \bs{k}_2, \bs{k}} 
\left[
(1 + N_{\bs{k}}^{\beta})(1 + N_{\bs{k}_2}^{\alpha}) N_{\bs{k}_3}^{\beta} N_{\bs{k}_4}^{\beta} 
- N_{\bs{k}}^{\beta} N_{\bs{k}_2}^{\alpha} (1 + N_{\bs{k}_3}^{\beta})(1 + N_{\bs{k}_4}^{\beta})
\right]
\bigg\}
\end{aligned}
\end{equation}

Now let us express collision term with 4-spin interaction. The first term represents decay of one $\beta$ magnon into three $\alpha$ magnons, while the second term is a decay of one $\beta$ magnon and one $\alpha$ magnon into two $\alpha$ magnons.

\begin{equation}
\begin{aligned}
I^{\textrm{4S}}_{\alpha} ={} &
\left( \frac{A_{\textrm{D}}}{32(S^{\textrm{A}} S^{\textrm{B}})^2} \right)^2 
\frac{1}{V_D^{2}} 
\int \frac{d^2\bs{k}_2}{(2\pi)^2} \frac{d^2\bs{k}_3}{(2\pi)^2} \frac{d^2\bs{k}_4}{(2\pi)^2} 
\, |\tilde{U}_{\bs{k} \bs{k}_2 \bs{k}_3 \bs{k}_4}|^2 \times \delta(\bs{k} + \bs{k}_2 + \bs{k}_3 + \bs{k}_4) \, 
\delta(\omega_{\bs{k}}^{\alpha} + \omega_{\bs{k}_2}^{\alpha} + \omega_{\bs{k}_3}^{\alpha} - \omega_{\bs{k}_4}^{\beta}) \\[4pt]
& \times \left[
(1 + N_{\bs{k}}^{\alpha})(1 + N_{\bs{k}_2}^{\alpha})(1 + N_{\bs{k}_3}^{\alpha}) N_{\bs{k}_4}^{\beta}
- N_{\bs{k}}^{\alpha} N_{\bs{k}_2}^{\alpha} N_{\bs{k}_3}^{\alpha}(1 + N_{\bs{k}_4}^{\beta})
\right]
\end{aligned}
\end{equation}

For deviation from the equilibrium $N_{\bs{k}}^{\alpha}  = N_{\bs{k}}^{\alpha,eq} + \delta N_{\bs{k}}^{\alpha} $, $N_{\bs{k}}^{\beta}  = N_{\bs{k}}^{\beta,eq} + N_{\bs{k}=0}^{\beta}(t) $ the collision terms becomes
\begin{equation}
\begin{aligned}
I_{\alpha\alpha} \approx  - \delta N_{\bs{k}}^{\alpha}/\tau_{\bs{k}}^{\textrm{ex}}.
\end{aligned}
\end{equation}
\begin{equation}
\begin{aligned}
I_{4S} \approx  N_{\bs{k}=0}^{\beta}(t)/\tau_{\bs{k}}^{\textrm{4S}}.
\end{aligned}
\end{equation}
with times
\begin{equation}
\begin{aligned}
\frac{1}{\tau_{\bs{k}}^{\textrm{ex}}} ={} &
\left( \frac{Jz}{4N} \right)^2 \frac{1}{V_D^2} 
\int \frac{d^2\bs{k}_2}{(2\pi)^2} \frac{d^2\bs{k}_3}{(2\pi)^2} \frac{d^2\bs{k}_4}{(2\pi)^2}
\, \left| V^{(1)}_{\bs{k} \bs{k}_2 \bs{k}_3 \bs{k}_4} \right|^2   \left[
N_{\bs{k}_2}^{\alpha,\textrm{eq}} (1 + N_{\bs{k}_3}^{\alpha,\textrm{eq}})(1 + N_{\bs{k}_4}^{\alpha,\textrm{eq}})
- (1 + N_{\bs{k}_2}^{\alpha,\textrm{eq}}) N_{\bs{k}_3}^{\alpha,\textrm{eq}} N_{\bs{k}_4}^{\alpha,\textrm{eq}}
\right] \\[4pt]
& \times \delta(\omega_{\bs{k}}^{\alpha} + \omega_{\bs{k}_2}^{\alpha} - \omega_{\bs{k}_3}^{\alpha} - \omega_{\bs{k}_4}^{\alpha})
\, \delta(\bs{k} + \bs{k}_2 - \bs{k}_3 - \bs{k}_4)
\end{aligned}
\end{equation}
\begin{equation}
\begin{aligned}
\frac{1}{\tau_{\bs{k}}^{\textrm{4S}}} = & \left( \frac{A_{\textrm{D}}}{32(S^{\textrm{A}} S^{\textrm{B}})^{3/2}} \right)^2 \frac{1}{V_D} 
\int \frac{d^2 \bs{k}_2}{(2\pi)^2} \frac{d^2 \bs{k}_3}{(2\pi)^2} 
|\tilde{U}_{\bs{k} \bs{k}_2 \bs{k}_3 0}|^2  \left[ (1 + N_{\bs{k}}^{\alpha,\textrm{eq}}) (1 + N_{\bs{k}_2}^{\alpha,\textrm{eq}}) (1 + N_{\bs{k}_3}^{\alpha,\textrm{eq}}) - N_{\bs{k}}^{\alpha,\textrm{eq}} N_{\bs{k}_2}^{\alpha,\textrm{eq}} N_{\bs{k}_3}^{\alpha,\textrm{eq}} \right] \\
& \cdot \delta \left( \omega_{\bs{k}}^{\alpha} + \omega_{\bs{k}_2}^{\alpha} + \omega_{\bs{k}_3}^{\alpha} - \Delta \right) 
\delta \left( \bs{k} + \bs{k}_2 + \bs{k}_3 \right)
\end{aligned}
\end{equation}

\section{Life time of magnon}
Life time of $\beta$-magnon can be computed as $1/\tau _{\beta}=\operatorname{Im}\Sigma^{R}(0,\Delta)/2$, where retarted self-energy is

\begin{equation}
\begin{aligned}
\Sigma(\Delta,0) &= -\left(\frac{A_{\textrm{D}}}{32(S^{\textrm{A}}S^{\textrm{B}})^{3/2}}\right)^{2}\frac{1}{V_{D}^{2}} \int \frac{d\omega_{2}}{2\pi} \frac{d\omega_{3}}{2\pi} \frac{d\omega_{4}}{2\pi} \frac{d^{2}k_{2}}{(2\pi)^{2}} \frac{d^{2}k_{3}}{(2\pi)^{2}} \frac{d^{2}k_{4}}{(2\pi)^{2}} \\
&\quad \delta(\boldsymbol{k}_{2}+\boldsymbol{k}_{3}+\boldsymbol{k}_{4}) \delta(\Delta-\omega_{2}-\omega_{3}-\omega_{4}) |\tilde{U}_{0,\boldsymbol{k}_{2},\boldsymbol{k}_{3},\boldsymbol{k}_{4}}|^{2} D(\boldsymbol{k}_{2},\omega_{2}) D(\boldsymbol{k}_{3},\omega_{3}) D(\boldsymbol{k}_{4},\omega_{4})
\end{aligned}
\end{equation}

Using optical theorem, we can compute imaginary part of self-energy $\operatorname{Im}\Sigma(0,\Delta)$  substituting Green function $D(\bs{k},\omega) = (\omega - \omega^{\alpha}_{\bs{k}}+i0)^{-1}$  by imaginary part of Green functions $D(\boldsymbol{k},\omega)\rightarrow \operatorname{Im} D(\boldsymbol{k},\omega)$, therefore we have
\be
\operatorname{Im}\Sigma(\Delta,0)=(\frac{A_{\textrm{D}}}{32(S^{\textrm{A}}S^{\textrm{B}})^{3/2}})^{2}\frac{1}{V_{D}^{2}}\int\frac{d^{2}k_{2}}{(2\pi)^{2}}\frac{d^{2}k_{3}}{(2\pi)^{2}} 
 \frac{d^{2}k_{4}}{(2\pi)^{2}}|\tilde{U}_{0,\boldsymbol{k}_{2},\boldsymbol{k}_{3},\boldsymbol{k}_{{4}}}|^{2}\delta(\Delta-\omega_{\boldsymbol{k}_{2}}-\omega_{\boldsymbol{k}_{3}}-\omega_{\boldsymbol{k}_{4}})\delta(\boldsymbol{k}_{2}+\boldsymbol{k}_{3}+\boldsymbol{k}_{4})
\e
with $|\tilde{U}_{0,\boldsymbol{k}_{2},\boldsymbol{k}_{3},\boldsymbol{k}_{4}}|^{2}=3\frac{(S^{\textrm{A}})^{2}+(S^{\textrm{B}})^{2}}{(1-\frac{S^{B}}{S^{A}})^{2}}(k_{2}^{2}+k_{3}^{2}+k_{4}^{2})(\frac{3}{2}a)^{2}$ and $\boldsymbol{q}=\frac{\boldsymbol{k}_{2}+\boldsymbol{k}_{3}}{2},\boldsymbol{p}=\frac{\boldsymbol{k}_{2}-\boldsymbol{k}_{3}}{2}$ we have 
\be
\operatorname{Im} \Sigma(0, \Delta) =
\frac{1}{8}
\left( \frac{A_{\textrm{D}}}{32 (S^{\textrm{A}} S^{\textrm{B}})^{3/2}} \right)^{2}
\frac{( (S^{\textrm{A}})^2 + (S^{\textrm{B}})^2 )}{\left( 1 - \frac{S^{\textrm{B}}}{S^{\textrm{A}}} \right)^{2}}
\cdot \frac{27}{4 V_D}
\int_0^\infty q \, dq \int_0^\infty p \, dp \;
\delta\left( \Delta - \frac{J S^{\textrm{A}} S^{\textrm{B}}}{2 (S^{\textrm{A}} - S^{\textrm{B}}) } (3q^2 + p^2)a^2 \right)
\e

As, a result we have 
\be
\frac{1}{\tau_{\beta}}=\frac{1}{2}\operatorname{Im}\Sigma(\Delta,0)=C\frac{A_{D}^{2}}{J}
\e
where

\be
C = \frac{9}{2^{17}\pi^{3}}\frac{((S^{\textrm{A}})^{2}+(S^{\textrm{B}})^{2})(S^{\textrm{A}}-S^{\textrm{B}})^{3}}{(S^{\textrm{A}})^{4}(S^{\textrm{B}})^{6}}
\e

\end{document}